\documentclass{aa}
\usepackage{graphicx}

\def\deg{$^\circ$}
\newcommand{\cmthree}{~cm$^{-3}$~}
\newcommand{\cmsix}{~cm$^{-6}$}

\newcommand{\kms}{km s$^{-1}$}
\newcommand{\etal}{et al.~}
\newcommand{\be}{\begin{equation}}
\newcommand{\ee}{\end{equation}}
\newcommand{\bea}{\begin{eqnarray}}
\newcommand{\eea}{\end{eqnarray}}
\newcommand{\HII}{\hbox{H\,{\sc ii}}~}
\newcommand{\CII}{\hbox{C\,{\sc ii}}~}
\newcommand{\FIRCII}{\hbox{[C\,{\sc ii}}]~158$\mu$m~}

\newcommand{\HI}{\hbox{H\,{\sc i}}~}
\newcommand{\CO}{$^{12}$CO~}
\newcommand{\lv}{\hbox{\em l-v}~}

\newcommand{\RGC}{R$_{\mbox{\sc GC}}$~}
\newcommand{\RO}{R$_\odot$}
\newcommand{\thetaz}{$\theta_0$}

\newcommand{\ie}{i.e.~}

\begin{document}

\title{Carbon Recombination Lines near 327 MHz }

\subtitle{I: ``Diffuse'' \CII regions in the Galactic Disk }

\author{D. Anish Roshi \inst{1,2} \thanks{ On leave from 
       National Centre for Radio Astrophysics,
       Tata Institute of Fundamental Research, Pune, India
       } 
      \and N. G. Kantharia \inst{2} \and K. R. Anantharamaiah \inst{3}\thanks{
      We regret to announce that our friend and collaborator,
      K. R. Anantharamaiah passed away on 29 October, 2001 before this
      work could be completed } }

\institute{
       National Radio Astronomy Observatory,
       Green Bank, WV 24944, USA. 
       \and National Centre for Radio Astrophysics, 
       Tata Institute of Fundamental Research, Pune, India 
       \and Raman Research Institute, Bangalore, India  }

\offprints{D. Anish Roshi \email{aroshi@nrao.edu}}

\date{Received  /Accepted }

\abstract{
In earlier papers (Roshi \& Anantharamaiah \cite{ra00},
\cite{ra01a}), we presented extensive surveys (angular resolution --
2\deg\ $\times$ 2\deg\ \& 2\deg\ $\times$ 6\arcmin)  of radio
recombination lines (RRLs) near 327 MHz in the longitude range $l = $
332\deg\ $\rightarrow$ 89\deg\ using the Ooty Radio Telescope.  These
surveys have detected carbon lines mostly between $l = $ 358\deg
$\rightarrow$ 20\deg\ and in a few positions at other longitudes.  This
paper presents the observed carbon line parameters in the high-resolution
survey and a study of the galactic distribution and angular extent of the
line emission observed in the surveys.  The carbon lines detected in the
surveys arise in ``diffuse'' \CII regions.  The \lv diagram and radial
distribution constructed from our carbon line data shows similarity with
that obtained from hydrogen recombination lines at 3 cm from \HII regions
indicating that the distribution of the diffuse \CII regions in the inner
Galaxy resembles the distribution of the star-forming regions.  We
estimated the \FIRCII emission from diffuse \CII regions and find
that upto 95 \% of the total observed \FIRCII emission can arise in diffuse
\CII regions if the temperature of the latter $\sim 80$ K. Our
high-resolution survey data shows that the carbon line emitting regions
have structures on angular scale $\sim$ 6\arcmin.  Analysis of the
dual-resolution observations toward a 2\deg\ wide field centered at $l = $
13\deg.9 and toward the longitude range $l = $ 1\deg.75 to 6\deg.75 shows
the presence of narrow ($\Delta V \le $ 15 \kms) carbon line emitting
regions extending over several degrees in $l$ and $b$.  The physical size
perpendicular to the line-of-sight of an individual diffuse \CII region in
these directions is $>$ 200 pc.

}

\titlerunning{Diffuse \CII regions in the Galactic Disk}
\authorrunning{D. Anish Roshi et al}
\maketitle

\section{Introduction}

Radio recombination lines (RRLs) of hydrogen, helium and carbon 
have been unambiguously identified in the spectra obtained 
toward \HII regions (see review by Roelfsema \& Goss \cite{rg92}).  The
hydrogen and helium recombination lines mostly originate in hot 
($T_e \sim$ 5000 -- 10000 K) regions ionized by photons of energy
$\ge$ 13.6 eV.  Since the ionization potential of carbon is 11.4 eV,
low energy photons (11.4 eV $ \le E < $ 13.6 eV) that escape from 
\HII regions can ionize gas phase carbon atoms outside the hot regions.
Thus ionized carbon regions can exist in dense 
(hydrogen nucleus density $n_0 \sim$ 10$^{5}$ \cmthree)
photo-dissociation regions (PDRs) adjacent to 
\HII regions or in the neutral components (\HI or molecular) of the interstellar
medium (ISM).
 Tielens and Hollenbach (\cite{th85}) define PDRs as 
regions where the heating or/and chemistry of the predominantly neutral gas is governed
by the FUV (6--13.6 eV) photons.  Since the FUV photons are omnipresent,
the PDRs, by definition encompass a substantial fraction of atomic gas
in a galaxy (Hollenbach \& Tielens \cite{ht97} and references therein). 
The dense PDRs (Tielens \& Hollenbach \cite{th85}) are located at
the interface of molecular clouds and \HII regions 
whereas the low-density ($n_0 \sim 10^3$ \cmthree) PDRs (Hollenbach, Takahashi \& Tielens \cite{htt91}) 
are located in the diffuse interstellar gas; 
the ambient FUV flux sufficing to control its chemistry and heating.
The ionized carbon regions in the dense PDRs are referred to
as ``classical'' \CII regions. 
These \CII regions are observationally identified 
by the narrow ( 1 -- 10 \kms) emission lines of carbon at frequencies $>$ 1 GHz.
Several studies have been made to understand and
model the line emission from such regions (eg. Garay \etal \cite{gar98}, 
Wyrowski \etal \cite{wetal00}).  
These regions are not accessible to low frequency RRLs due to the
increased pressure broadening ($ \propto \nu^{-8.2/3}$; Shaver \cite{s75}) 
and increased free-free continuum optical
depths ($\tau \propto \nu^{-2}$). 
The second class of \CII regions, referred to as ``diffuse'' \CII regions,
coexists with the diffuse neutral component of the ISM. 
The emission measures 
of these regions are fairly low ($< 0.1$ \cmsix pc;  Kantharia, Anantharamaiah \& Payne \cite{kap98})
and hence these regions are observable in low-frequency RRLs of carbon as either
absorption lines or emission lines due to stimulated emission from inverted populations.
The diffuse \CII regions,
observed in carbon lines at frequencies $ < 1 $ GHz, are the focus of this paper.

The diffuse \CII region located in the Perseus arm toward the strong radio
continuum source, Cas~A has been extensively studied using low frequency recombination
lines of carbon.  In fact, most of our knowledge on this class of \CII regions
has come from these observations.
Konovalenko \& Sodin (\cite{ks80}) were the first to observe a low-frequency
(26.3 MHz) absorption line toward Cas A, which was later correctly
identified as the 631$\alpha$ recombination line of carbon by
Blake, Crutcher \& Watson (\cite{bcw80}).  Since then, several
recombination line observations spanning over 14 to 1400 MHz have been made
toward this direction (Kantharia \etal \cite{kap98} 
and references therein).  The predicted smooth transition of 
carbon lines in absorption at 
frequencies below 115 MHz to lines in emission at frequencies above 200 MHz 
has been demonstrated toward this direction 
(Payne, Anantharamaiah \& Erickson \cite{pae89}).  The extensive RRL data collected
toward Cas~A has
been used in modeling the line-forming gas.  The models show that the 
carbon RRLs originate in small, relatively cool tenuous regions 
($T_e$ = 35--75 K, $n_e$ = 0.05--0.1 \cmthree, size $\sim 2$ pc;
Payne, Anantharamaiah \& Erickson \cite{pae94}) 
of the ISM. 
Comparison of the distribution of carbon RRLs near 327 MHz observed with the
VLA (2\arcmin.7 $\times$ 2\arcmin.4) toward Cas A with \HI absorption in the
same direction suggests that the carbon line-forming region likely
coexists with the cold, diffuse \HI component of the ISM 
(Anantharamaiah \etal \cite{aepk94}).  

In addition to the region toward Cas~A,  
the distribution of the diffuse \CII regions in the Galaxy has also been studied
to some extent.  Surveys have been conducted 
near 76 MHz ($n \sim 441$) with the Parkes 64m telescope (Erickson, 
McConnell \& Anantharamaiah \cite{ema95}) and near 35 MHz ($n \sim 580$)
with the Gauribidanur telescope (Kantharia \& Anantharamaiah \cite{ka01}) to
search for carbon recombination lines, mostly in the inner part of the Galaxy. 
These observations have succeeded in detecting carbon RRLs in absorption 
from several directions in the galactic plane with longitudes ranging from 
$l = $ 340\deg\ $\rightarrow$ 20\deg. The diffuse \CII regions appear to 
be fairly widespread in the inner part of our Galaxy.   
Observations away from the Galactic
plane have shown the region to be several degrees wide in galactic latitude. 
The positions with detections near 35 MHz were observed near 327 MHz 
using the Ooty Radio Telescope by Kantharia \& Anantharamaiah (\cite{ka01})
and the emission counterparts of the carbon absorption lines were detected.  
Combining their observations with all other existing carbon RRL observations, 
they modeled the line emission at different positions in the galactic plane. 
While models with physical properties 
similar to those obtained in the direction of Cas A can fit
the observed data, the possibility of carbon lines originating
in regions with temperature $\le$ 20 K cannot be ruled out
(Kantharia \& Anantharamaiah \cite{ka01}).  If the temperature of
the diffuse \CII regions is found to be low, then
these regions could even be associated with the molecular component of the ISM
(Konovalenko \cite{k84}, Golynkin \& Konovalenko \cite{gk90}, Sorochekov \cite{s96},
Kantharia \& Anantharamaiah \cite{ka01}).  These low temperature
regions may be low-density PDRs (Hollenbach, Takahashi, Tielens \cite{htt91}) 
formed on surfaces of molecular clouds due to
ionization from background FUV radiation.  
Although some modeling of these diffuse \CII regions using low-frequency
carbon RRLs has been done, 
a wide range of parameter space has been found to fit the existing
observations.  The physical properties, distribution and association of these 
regions with other components of the ISM requires more investigation.
In addition to carbon RRLs, ionized carbon is also traced by the \FIRCII line.
The $158\mu$ line emission from the Galaxy has been mapped by
Bennett \etal (\cite{ben94}) and  Nakagawa \etal (\cite{naka98}).  They find that
the \FIRCII emission consists of compact emission regions associated
with compact \HII regions (Nakagawa \etal \cite{naka98}) and a diffuse emission whose
origin is not very clear.
Since both the fine-structure line and the carbon RRLs require ionized carbon
regions, it is possible that the two can arise from similar regions.
Kantharia \& Anantharamaiah (\cite{ka01}) tried to compare the carbon
lines near 35 MHz with the \FIRCII emission but they did not derive any conclusive results.
Hence, no detailed comparative study of the radio and FIR line emission
of carbon from diffuse \CII regions exists.  In this paper, we also attempt
a discussion on these two tracers of ionized carbon regions.

Extensive surveys of recombination lines near 327 MHz 
have been made with the primary objective to study the  
low-density ionized gas in the Galaxy by observing low-frequency hydrogen
RRLs from this gas (Roshi \& Anantharamaiah \cite{ra00}; hereafter Paper I; 
Roshi \& Anantharamaiah \cite{ra01a}; hereafter Paper II; 
Roshi \& Anantharamaiah \cite{ra01b}).  Since the velocity coverage
of these surveys was sufficient to allow detection of carbon RRLs, which are
separated from the hydrogen line by $\sim -150$ \kms\ , 
the surveys have succeeded in detecting carbon features toward several
positions in the galactic plane.  The surveys have data with two 
different angular resolutions
obtained by using the Ooty Radio Telescope in two different 
operating modes (see Paper I \& II).  
The carbon line data obtained from the higher angular resolution observation
(2\deg\ $\times$ 6\arcmin\ ) are presented in this paper (see Paper II for spectra) 
and those obtained in the lower resolution (2\deg\ $\times$ 2\deg\ ) 
survey were presented in Paper I. 
In this paper, we present a study of the distribution and angular extent of the carbon
line forming region in the galactic plane by making use of the 
carbon RRLs detected in the 327 MHz surveys. 
Interestingly, in several directions the carbon line
emission observed in the surveys seems to be associated with 
\HI\ self-absorption features, which will be discussed in 
Roshi, Kantharia \& Anantharamaiah (\cite{rka02}).
 
A summary of the observations  
and basic results are presented in Section~\ref{sec:obs}. 
Section~\ref{sec:dis} discusses
the distribution of the diffuse \CII regions in the galactic disk
and compares it with the distribution of other components of the ISM.
Section~\ref{finestructure}
discusses the possibility of a common origin of the carbon RRL and the
diffuse \FIRCII line emission.
The latitude extent of carbon line emission is discussed in
Section~\ref{sec:lat}.  The higher resolution observations are
used to study the angular extent of the carbon line emitting region,
which is discussed in Section~\ref{sec:ang}. Section~\ref{sec:sum} 
summarizes the paper.

\section{Summary of Observations and Basic Results}
\label{sec:obs}

The RRL surveys, which were described in detail in Papers I \& II,
were made using the Ooty Radio Telescope (ORT).  ORT is a 530m $\times$ 30m
parabolic cylinder operating at a nominal center frequency of 327 MHz
(Swarup \etal\ \cite{setal71}). 
The observations were made with two different angular resolutions
-- (a)  2\deg\ $\times$ 2\deg\ (low resolution mode) and (b) 2\deg\ $\times$ 6\arcmin\ 
(high resolution mode).  The high resolution mode is
obtained by using all the 22 `modules' of the ORT, which together form a telescope
of size 530 m $\times$ 30 m, and the low resolution mode
is obtained by using only a single `module' of the ORT, which effectively is a telescope
of size 24 m $\times$ 30 m.
The RRL transitions from principal quantum numbers $n$ = 270, 271, 272 and
273 and $\Delta n$ = 1 were simultaneously observed using a 
multi-line spectrometer (Roshi \cite{r99}).  The final spectrum is 
obtained by averaging all the four RRL transitions. 

\subsection{Low-resolution Survey}
\label{sec:lowres}

In the low-resolution survey (Paper I), 
51 positions were observed in the inner Galaxy: longitude range
$l$ = 332\deg\ to 0\deg\ to 89\deg\ and $b=0$\deg. The positions
were separated in longitude by
$\sim$2\deg\ $\times$ $\sec(\delta)$, $\delta$ being the declination.
Carbon RRLs were detected from almost all directions 
in the longitude range $l = $ 358\deg\ $\rightarrow$ 20\deg\
and also from a few positions in the longitude range $l = $ 20\deg\ to 89\deg.
In the outer Galaxy (172\deg\ $<~l~<$ 252\deg) a total of 14 positions,
spaced by $\sim$ 5\deg -- 7\deg\ in longitude, were observed.  
However, no carbon RRLs were detected in this longitude range. 
At two specific longitudes
in the inner Galaxy ($l$ = 0\deg.0 \& 13\deg.9), spectra were taken in
steps of 1\deg\ up to $b = \pm 4$\deg\ to study the latitude extent of
the carbon line emission. 
The observed spectra and line fit parameters were presented in Paper I. 

\subsection{High-resolution Survey}
\label{sec:higres}

In the high-resolution survey (Paper II), a set of seven 
fields which were 2\deg\ wide and two fields
which were 6\deg\ wide in longitude were observed with a 2\deg\ $\times$ 6\arcmin\ beam.  
The fields are designated as Field 1 to 9
and are centered at $l = $ 348\deg.0 (2\deg\ wide), 3\deg.4 (6\deg), 13\deg.9 (2\deg), 
25\deg.2 (2\deg), 27\deg.5 (2\deg), 36\deg.3 (6\deg),
45\deg.5 (2\deg), 56\deg.9 (2\deg) and 66\deg.2 (2\deg) respectively.
The ORT is an equatorially mounted telescope and the beam size
is 2\deg\ along right ascension. The orientation of the beam with respect to
galactic co-ordinates, therefore, changes as a 
function of galactic longitude. 
Carbon RRLs were detected toward several positions within the fields with $l <$ 40\deg,
whereas no lines were detected at any individual positions within the fields in the
longitude range $l = $ 40\deg\ to 85\deg.  The galactic coordinates of the positions
where carbon lines are detected and the parameters estimated from Gaussian fits to 
the line profiles are given in Table~\ref{tab:spectra}. 
Each spectrum was inspected by eye and the presence of a carbon line was determined.
If narrow ($\sim$ 1-2 channels) spectral features were present in addition to the carbon
line feature, we regarded the detection as tentative. The narrow spurious features were either due to
residual radio frequency interference or ``bad'' spectral channel values, which were
inferred from the channel weights as discussed in Paper I. 
However, if the width of the carbon line was several times ($\sim$ 10) 
larger than any spurious narrow features then we regarded them as real.
Since the peak line intensity to the RMS noise in the spectra is
about 3 to 4, care has been taken in fitting Gaussian components to the line profile. 
A second Gaussian component was fitted to only those spectra where the residuals
left after removing a single Gaussian component were inconsistent with the noise in
rest of the spectrum. 
The details of the high-resolution survey and the observed spectra
were presented in Paper II.

\subsection{Line Width}

\begin{figure}
 \resizebox{\hsize}{!}{\includegraphics{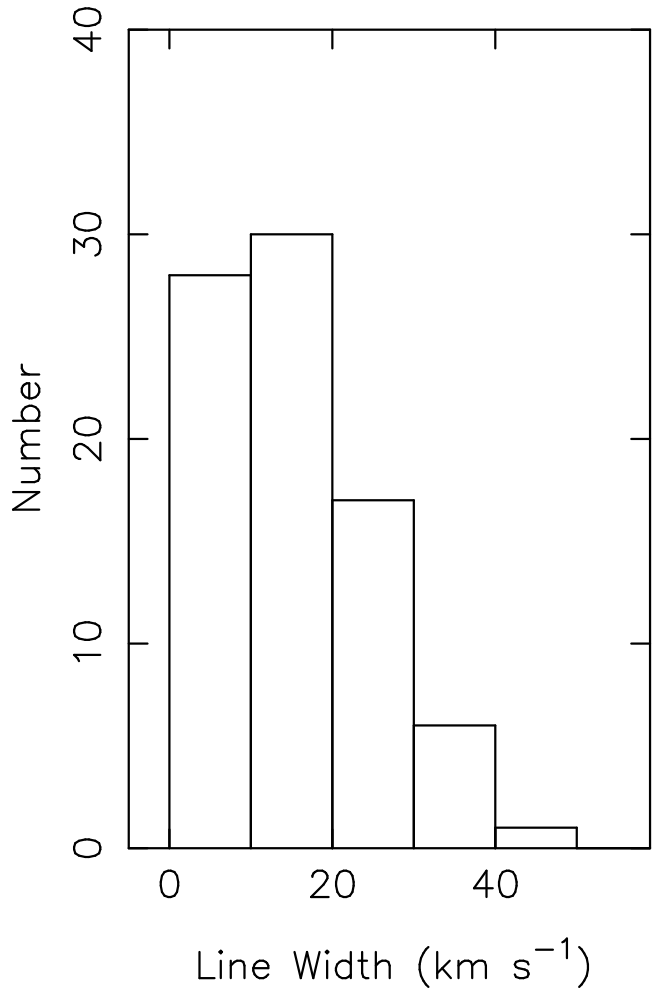}}
 \caption{Histogram of the observed carbon line widths
 in the high-resolution (2\deg\ $\times$ 6\arcmin\ ) survey data.}
 \label{fig:whis}
\end{figure}

If we assume that the physical properties of the carbon line emitting gas observed
in the galactic disk are similar to those derived for the gas
toward Cas A (e.g. $T_e = 75$K, $n_e=0.1$ \cmthree, from Payne \etal\ \cite{pae94})
and the galactic background radiation temperature $=700$ K (from Paper I),
then the total contribution from pressure, radiation and Doppler 
broadening at 325 MHz would be a negligible $\sim 0.6$ \kms. 
Comparing this with the observed line widths in the high-resolution data 
ranging from 4 to 48 \kms\ with a median
value of 14 \kms\ (Fig.~\ref{fig:whis}), it is clear that the 
cause of line broadening lies elsewhere.
Blending of carbon line features from different line forming regions 
within the coarse survey beam and turbulent motions within the cloud
are likely the cause of the broad lines.  To confirm this 
we have examined the data toward a few directions in the 
galactic plane in more detail and the results (see Section \ref{sec:ang} for details)
support our conclusion. 
Moreover, the median line width obtained from the low-resolution survey 
data is 17 \kms (Paper I), which
is somewhat larger than the value estimated from the higher resolution data ($\sim 14$ \kms). 
The larger median value is likely a result of line blending because of the
relatively larger beam width of the low-resolution survey.  
The line widths of carbon lines observed in the survey are typically a factor
of 2 to 5 larger than the typical line width of 
carbon lines observed at frequencies $>$ 1 GHz from ``classical'' \CII regions 
(eg. Roelfsema \& Goss \cite{rg92}). Since the lines at low and high frequencies
are believed to arise in distinct regions of the ISM, the difference is not surprising.  

\clearpage
%
%
\begin{table}
\caption{Summary of the carbon RRL observations from the high resolution survey}
\label{tab:spectra}
\begin{tabular}{rrrrrccr}
\hline 
\hline
\multicolumn{1}{c}{$ l $}& \multicolumn{1}{c}{$b$} & \multicolumn{1}{c}{$ T_L/T_{sys}\footnotemark[1]$} & 
\multicolumn{1}{c}{$\Delta  V$} & \multicolumn{1}{c}{$V_{\rm{LSR}}$} & 
\multicolumn{1}{c}{$V_{res}\footnotemark[2]$} & RMS \footnotemark[3] & $t_{int}$ \\
\multicolumn{1}{c}{$^o$}&\multicolumn{1}{c}{$^o$}&\multicolumn{1}{c}{$\times$ 10$^{3}$}&\multicolumn{1}{c}{\kms} &\multicolumn{1}{c}{\kms}     & \kms & $\times$ 10$^{3}$ & \multicolumn{1}{c}{hrs} \\
\hline
\multicolumn{8}{c}{\bf Field 2a} \\
\hline
    0.52 & $+$0.03 & 0.64(0.08) & 22.4(3.2) &    1.7(1.4) & 1.8 & 0.20 & 12.8   \\
    0.67 & $-$0.00 & 0.66(0.18) &  4.3(1.3) &    6.7(0.6) & 1.8 & 0.19 & 11.2   \\
         &         & 0.37(0.15) &  5.8(2.7) & $-$11.1(1.2) & 1.8 & 0.19 & 11.2   \\
    0.75 & $+$0.05 & 0.54(0.11) &  9.8(2.3) &   18.9(1.0) & 3.4 & 0.13 & 11.6   \\
     &  & 0.41(0.09) & 14.8(3.7) & $-$7.2(1.6) & 3.4 & 0.13 & 11.6   \\
    0.84 & $+$0.10 & 0.69(0.12) & 15.0(3.0) &    6.3(1.3) & 3.4 & 0.18 &  9.4   \\
    0.92 & $+$0.16 & 0.83(0.22) &  4.0(1.2) &    6.5(0.5) & 1.8 & 0.23 & 12.6   \\
\hline
\multicolumn{8}{c}{\bf Field 2b (G2.3+0.0)} \\
\hline
    1.21 & $+$0.07 & 0.32(0.1)\footnotemark[4] & 11.1(4.2) &    2.6(1.7) & 3.4 & 0.13 & 10.3   \\
     &  & 0.33(0.11)\footnotemark[4] &  9.2(3.6) & $-$17.8(1.5) & 3.4 & 0.13 & 10.3   \\
    1.29 & $+$0.13 & 0.58(0.08) & 25.5(4.2) &    3.6(1.8) & 2.1 & 0.21 &  9.9   \\
    1.38 & $+$0.18 & 0.55(0.08) & 30.1(5.3) &    1.6(2.2) & 3.4 & 0.17 & 10.1   \\
    1.83 & $+$0.20 & 0.63(0.15) &  7.3(2.0) &    1.9(0.9) & 2.1 & 0.20 & 11.3   \\
    2.29 & $+$0.21 & 0.47(0.09) & 14.3(3.1) &   10.3(1.3) & 2.1 & 0.17 & 14.9   \\
    2.54 & $-$0.03 & 0.53(0.13) & 14.3(4.0) &    8.9(1.7) & 2.1 & 0.24 &  8.1   \\
    2.63 & $+$0.02 & 0.55(0.12) & 10.3(2.6) &    4.0(1.1) & 2.1 & 0.19 & 11.2   \\
    2.78 & $-$0.03 & 0.34(0.08) & 18.0(4.6) &   11.1(1.9) & 2.1 & 0.16 & 11.2   \\
    2.86 & $+$0.02 & 0.32(0.07) & 28.7(7.9) &   10.6(3.3) & 2.1 & 0.20 & 12.5   \\
    3.01 & $-$0.03 & 0.44(0.08) & 26.3(5.3) &    9.3(2.2) & 2.1 & 0.19 & 11.2   \\
    3.09 & $+$0.02 & 0.49(0.12) & 12.8(3.6) &    5.8(1.5) & 2.1 & 0.21 & 11.7   \\
    3.33 & $+$0.02 & 0.46(0.1) & 14.0(3.7) &    6.9(1.5) & 2.1 & 0.19 & 11.5   \\
\hline
\multicolumn{8}{c}{\bf Field 2c (G4.7+0.0)} \\
\hline
    3.56 & $+$0.02 & 0.33(0.08) & 19.5(5.3) &   10.0(2.2) & 3.4 & 0.13 & 10.4   \\
    3.79 & $+$0.02 & 0.45(0.12) & 11.8(3.6) &    9.8(1.5) & 3.4 & 0.16 & 11.6   \\
    3.94 & $-$0.03 & 0.66(0.15) &  8.8(2.3) &   11.6(1.0) & 2.1 & 0.22 &  8.4   \\
    4.26 & $+$0.02 & 0.72(0.15) &  7.8(1.9) &    5.8(0.8) & 2.1 & 0.21 & 12.2   \\
    4.49 & $+$0.02 & 0.40(0.12)\footnotemark[4] & 14.4(5.0) &    9.2(2.1) & 3.4 & 0.17 & 11.6   \\
    4.64 & $-$0.03 & 0.27(0.08)\footnotemark[4] & 22.8(7.9) &    8.0(3.3) & 3.4 & 0.15 &  9.6   \\
    4.72 & $+$0.02 & 0.39(0.09) & 28.2(7.5) &   10.6(3.2) & 3.4 & 0.18 & 11.5   \\
    4.87 & $-$0.03 & 0.79(0.18)\footnotemark[4] &  7.0(1.9) &   11.9(0.8) & 1.8 & 0.25 &  8.3   \\
     &  & 0.75(0.22)\footnotemark[4] &  4.9(1.6) & $-$35.5(0.7) & 1.8 & 0.25 &  8.3   \\
    4.95 & $+$0.02 & 0.64(0.14) & 11.4(2.8) &    9.0(1.2) & 2.1 & 0.23 &  8.5   \\
    5.19 & $+$0.02 & 0.78(0.15) &  8.1(1.8) &    8.4(0.8) & 1.8 & 0.23 &  9.0   \\
    5.33 & $-$0.03 & 0.54(0.07) & 22.6(3.6) &   12.8(1.5) & 2.1 & 0.17 & 10.2   \\
    5.42 & $+$0.02 & 0.49(0.09) & 20.1(4.0) &    9.1(1.7) & 2.1 & 0.19 & 10.6   \\
    5.56 & $-$0.03 & 0.71(0.19)\footnotemark[4] &  4.3(1.4) &    7.5(0.6) & 2.1 & 0.20 & 10.0   \\
    5.65 & $+$0.02 & 0.79(0.15) &  7.1(1.5) &    5.6(0.6) & 2.1 & 0.19 & 12.3   \\
    5.80 & $-$0.03 & 0.80(0.15) &  9.1(1.9) &    5.2(0.8) & 2.1 & 0.22 &  7.4   \\
     &  & 0.72(0.19) &  5.5(1.7) &   18.4(0.8) & 2.1 & 0.22 &  7.4   \\
    5.88 & $+$0.02 & 0.51(0.13) & 14.0(4.0) &   13.7(1.7) & 2.1 & 0.23 &  8.6   \\
    6.02 & $-$0.02 & 0.43(0.14)\footnotemark[4] & 20.3(7.7) &    6.0(3.2) & 3.4 & 0.24 &  4.2   \\
    6.25 & $-$0.02 & 0.67(0.14)\footnotemark[4] & 16.0(3.9) &    0.6(1.6) & 2.1 & 0.27 &  5.6   \\
    6.72 & $-$0.02 & 0.53(0.12) &  9.7(2.6) &    6.1(1.1) & 2.1 & 0.19 & 10.0   \\
    6.80 & $+$0.03 & 0.58(0.12) &  8.6(2.0) &    7.7(0.9) & 2.1 & 0.17 & 11.4   \\
\hline
\end{tabular}
\end{table}
\clearpage
\begin{table}
\begin{tabular}{rrrrrccr}
\hline
\multicolumn{8}{c}{\bf Field 3 (G13.9+0.0)}\\
\hline
   13.04 & $-$0.46 & 0.46(0.07) & 36.0(6.6) &   40.6(2.8) & 2.1 & 0.21 &  7.9   \\
   13.13 & $-$0.41 & 0.46(0.14) &  7.0(2.5) &   52.1(1.0) & 2.1 & 0.18 & 10.6   \\
    &  & 0.89(0.11) & 10.8(1.6) &   35.2(0.7) & 2.1 & 0.18 & 10.6   \\
    &  & 0.64(0.11) & 12.5(2.4) &   15.7(1.0) & 2.1 & 0.18 & 10.6   \\
   13.22 & $-$0.36 & 0.41(0.08) & 37.2(8.1) &   39.9(3.4) & 1.8 & 0.24 &  8.2   \\
    &  & 0.61(0.16) &  8.0(2.5) &   18.4(1.1) & 1.8 & 0.24 &  8.2   \\
   13.30 & $-$0.31 & 0.51(0.1) & 24.8(5.8) &   42.3(2.4) & 1.8 & 0.26 &  8.4   \\
    &  & 0.99(0.17) &  8.6(1.7) &   19.8(0.7) & 1.8 & 0.26 &  8.4   \\
   13.39 & $-$0.26 & 0.44(0.12) & 14.4(4.6) &   11.8(2.0) & 2.1 & 0.23 &  8.4   \\
    &  & 0.33(0.09) & 29.8(8.8) &   39.5(3.7) & 2.1 & 0.23 &  8.4   \\
   13.48 & $-$0.22 & 0.38(0.11) & 28.6(9.7) &   44.0(4.1) & 3.4 & 0.23 & 11.2   \\
    &  & 0.50(0.15) & 16.5(5.6) &   14.5(2.4) & 3.4 & 0.23 & 11.2   \\
   13.57 & $-$0.17 & 0.77(0.18) &  5.4(1.5) &   19.3(0.6) & 2.1 & 0.21 &  9.3   \\
    &  & 0.43(0.07) & 38.9(7.0) &   33.1(3.0) & 2.1 & 0.21 &  9.3   \\
   13.65 & $-$0.12 & 0.41(0.09) & 20.8(5.4) &   24.8(2.3) & 2.1 & 0.20 &  9.1   \\
   13.74 & $-$0.07 & 0.57(0.17)\footnotemark[4] &  9.0(3.2) &   18.0(1.3) & 2.1 & 0.25 &  8.0   \\
   13.83 & $-$0.02 & 0.49(0.14)\footnotemark[4] & 10.1(3.3) &   44.5(1.4) & 2.1 & 0.21 &  8.8   \\
    &  & 0.66(0.19)\footnotemark[4] &  5.2(1.8)&   18.7(0.7) & 2.1 & 0.21 &  8.8   \\
   13.92 & $+$0.03 & 0.84(0.21)\footnotemark[4] &  6.5(1.9) &   19.2(0.8) & 2.1 & 0.26 &  6.3   \\
   14.09 & $+$0.12 & 0.45(0.05) & 47.3(6.3) &   34.3(2.7) & 2.1 & 0.18 & 10.0   \\
   14.18 & $+$0.17 & 0.35(0.09) & 19.1(5.7) &   39.8(2.4) & 2.1 & 0.19 & 10.9   \\
    &  & 0.61(0.13) &  8.6(2.2) &   19.6(0.9) & 2.1 & 0.19 & 10.9   \\
   14.36 & $+$0.26 & 0.58(0.06) & 29.3(3.6) &   43.6(1.5) & 2.1 & 0.16 & 11.7   \\
    & & 0.57(0.1) & 11.3(2.3) &   18.7(1.0) & 2.1 & 0.16 & 11.7   \\
   14.44 & $+$0.31 & 0.52(0.08) & 27.1(5.1) &   19.6(2.1) & 2.1 & 0.21 &  7.5   \\
   14.62 & $+$0.41 & 0.52(0.14) &  9.9(3.1) &   26.9(1.3) & 2.1 & 0.22 & 10.1   \\
   14.71 & $+$0.46 & 0.35(0.1) & 12.8(4.4) &   51.5(1.9) & 2.1 & 0.18 & 11.8   \\
    &  & 0.43(0.08) & 20.0(4.5) &   23.0(1.9) & 2.1 & 0.18 & 11.8   \\
\hline
\multicolumn{8}{c}{\bf Field 5 (G27.5+0.0)}\\
\hline
   27.06 & $-$0.20 & 0.41(0.09) & 19.4(5.1) &   58.2(2.2) & 3.4 & 0.16 &  8.0   \\
   28.04 & $+$0.31 & 0.66(0.15) &  6.6(1.7) &   79.5(0.7) & 1.8 & 0.20 & 15.0   \\
   28.13 & $+$0.35 & 0.34(0.07) & 20.7(5.1) &   73.2(2.2) & 3.4 & 0.13 & 18.6   \\
\hline
\multicolumn{8}{c}{\bf Field 6a (G34.2+0.0)}\\
\hline
   34.18 & $-$0.02 & 0.63(0.15)\footnotemark[4] &  8.1(2.3) &   44.4(1.0) & 2.1 & 0.22 &  9.7   \\
   34.27 & $+$0.03 & 0.36(0.08) & 15.2(3.9) &   49.2(1.6) & 2.1 & 0.15 & 13.1   \\
   34.85 & $-$0.02 & 0.55(0.08) & 31.3(5.6) &   43.8(2.3) & 2.1 & 0.23 &  8.9   \\
   34.94 & $+$0.03 & 0.45(0.1) & 18.1(4.7) &   41.0(2.0) & 2.1 & 0.21 & 11.0   \\
   35.08 & $-$0.02 & 0.38(0.1) & 17.4(5.0) &   51.6(2.1) & 3.4 & 0.15 &  8.9   \\
   35.17 & $+$0.03 & 0.51(0.08) & 26.6(5.1) &   50.2(2.2) & 3.4 & 0.17 & 11.9   \\
   35.31 & $-$0.02 & 0.23(0.07) & 19.1(7.0) &   66.4(2.9) & 7.6 & 0.82 & 11.1   \\
\hline
\multicolumn{8}{c}{\bf Field 6b (G36.5+0.0)}\\ 
\hline
   35.76 & $-$0.02 & 0.56(0.10) & 12.6(2.5) &   51.5(1.0) & 2.1 & 0.17 & 10.7   \\
   36.21 & $-$0.02 & 0.43(0.07) & 30.3(5.5) &   52.7(2.3) & 2.1 & 0.18 & 10.5   \\
\hline
\end{tabular}
\end{table}
\footnotetext[1]{The line intensities are given in units of $T_L/T_{sys}$,
where $T_L$ is the line antenna temperature and $T_{sys}$ is the
system temperature, which includes
sky, receiver and spillover temperature} 
\footnotetext[2]{The spectral resolution in \kms. }
\footnotetext[3]{RMS is in units of $T_L/T_{sys}$.} 
\footnotetext[4]{Tentative detection.}

\clearpage

\section{Distribution of the Carbon Line Forming Gas in the Galactic disk}
\label{sec:dis}

\subsection{Observed Longitudinal Variation of Integrated Line Intensity} 
\label{sec:lvsl}

\begin{figure}
 \resizebox{\hsize}{!}{\includegraphics{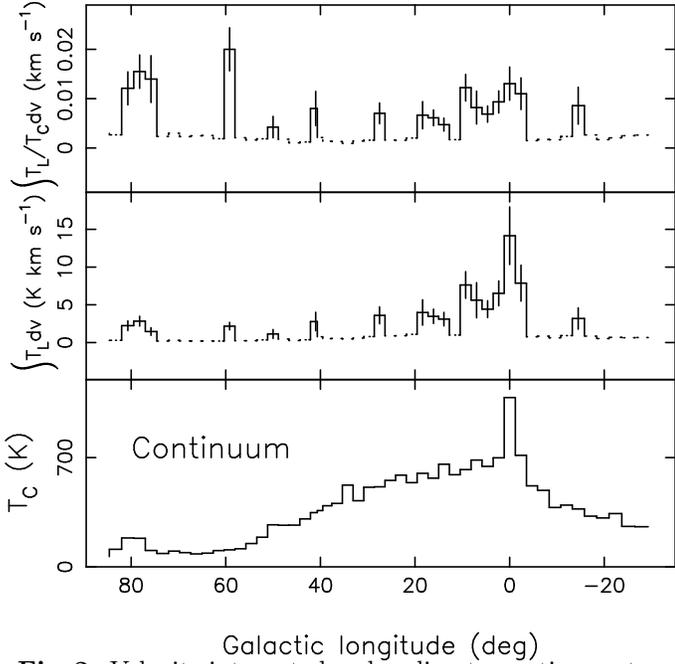}}
 \caption{ Velocity integrated carbon line-to-continuum temperature ratio (top panel),
 velocity-integrated carbon line temperature (middle panel) and
 the continuum emission (bottom panel) near 327 MHz,
 observed in the low-resolution (2\deg\ $\times$ 2\deg) survey, plotted as
 a function of galactic longitude.  The horizontal lines in the top and middle
 panels indicate the observed positions where carbon lines are detected.
 The vertical bar represents the 3 $\sigma$ uncertainty in the plotted parameters. 
 The dashed line in these plots indicate observed positions with no
 detections and give the upper limit on the quantities 
 plotted.  These limits were estimated from the RMS noise ($\sigma$)
 on the spectra and assuming a typical
 width for the carbon line as 17 \kms.  
 The continuum temperature plotted
 in the bottom panel is the measured antenna temperature corrected for the
 beam efficiency factor (0.65) and is same as $T_C$ discussed in Section 3.1
 under the assumptions stated in that section. }
 \label{fig:cltctl}
\end{figure}

The variation of velocity-integrated line-to-continuum temperature ratio,
velocity-integrated line strength of carbon RRLs and continuum temperature with
galactic longitude observed in the low-resolution survey is shown in
Fig.~\ref{fig:cltctl}.  
The line-to-continuum ratio for optically thin case is approximately given by 
(Shaver \cite{s75}),
\begin{equation}
\frac{T_L}{T_C} \sim -\tau_L + \frac{T_e \tau_L}{T_C \beta_n},
\label{eqn1}
\end{equation}
where $T_L$, $T_C$ and $T_e$ are the line brightness temperature, 
continuum brightness temperature and electron temperature respectively and 
$\tau_L$ is the non-LTE line optical depth. 
$\beta_n \sim 1 - \frac{kT_e}{h\nu} \frac{d~ln(b_n)}{dn}$ for 
$\Delta n $ = 1 transition. Here $k$ is the
Boltzmann's constant, $h$ is the Planck's constant, 
$\nu$ is the frequency of the line emission and $b_n$
is the departure coefficient. $\beta_n$ is a measure of the non-LTE 
effects on the level populations.   
For getting Eq. \ref{eqn1}, it is also assumed that the beam dilution is 
negligible and the measured
continuum temperature is approximately the background temperature at the 
location of the line-forming region. 
Since $-\beta_n \sim 50-100$ at levels near $n \sim 272$ for typical 
physical conditions of the line-forming region,
the line-to-continuum ratio is approximately the line optical depth 
even for the case where $T_e \sim T_C$.
If we assume that beam dilution is negligible and the background
temperature is the measured continuum
temperature, both of which might not be true for all longitudes, then we
can conclude that the top panel of Fig.~\ref{fig:cltctl} shows the
variation of velocity-integrated line optical
depth with longitude.  If the assumptions are not valid, then the plotted values will
be the apparent integrated optical depth which would be a lower limit on the 
actual integrated optical depth. 
The integrated optical depth (top panel in Fig.~\ref{fig:cltctl}) 
is constant ($\sim$ 0.01 \kms) 
within $3\sigma$ measurement errors for most of our detections.
From the top and middle panels, it is evident that line
emission is concentrated in the longitude range 358\deg\ $\rightarrow$
20\deg\ with a few detections at longitudes between 20\deg\ and 89\deg\ .  
The strongest integrated line strength and continuum are observed toward
the Galactic center.  This is a good case of stimulated
emission; the optical depth toward this region is not enhanced as
compared to the neighboring longitudes.  The bump near $l = $ 80\deg\
(see Fig.~\ref{fig:cltctl}) seen in all three distributions
is from the gas associated with the well-known Cygnus complex located in
the nearby Orion spiral arm.  
There are only a couple of detections in the fourth quadrant. 

After examining the observed distribution of the line emission with
longitude (Fig.~\ref{fig:cltctl}) and various factors at play, we conclude
that the paucity of detections at longitudes outside the range 358\deg $<
l <$ 20\deg\ may not be real but a result of one or more of the following
selection effects: 1)  reduced background radiation field leading to
reduced stimulated emission and hence weaker lines.  This, we believe is
the reason for fewer detections at longitudes $ > 20$ \deg.  Since the
intensity of carbon RRLs is amplified by the non-thermal background
continuum due to stimulated emission (Paper I), the gradual drop in the
non-thermal continuum with increasing longitudes might be partially
responsible for the drop in the line strengths and subsequently lesser
number of detections between longitudes 20\deg\ and 80\deg.  2) Beam
dilution within the large low resolution survey beam leading to reduced
line strengths and our sensitivity-limited sample failing to detect these
lines.  This is likely the dominant cause of non-detection of lines in the
fourth quadrant.  The ORT has an equatorial mount and electrical phasing
is used to point the telescope along the declination axis.  At longitudes
$l < 355$\deg\ due to a variety of reasons (eg. improper phasing) the
telescope sensitivity drops and also the beam size increases
(Roshi \cite{r99}). The drop in the continuum temperature at these
longitudes (Fig.~\ref{fig:cltctl}) is a result of this effect.  On
the other hand, negligible beam dilution effects could be one of the
reasons we detect the carbon lines from the Cygnus region ($l \sim
80$\deg\ ) located in the nearby Orion arm despite the background
radiation field being weaker than the regions between $l=20$\deg\ to $80$\deg\
and the presence of increased beam size as in negative 
longitudes.

The few positions where carbon lines were detected in the
longitude range 20\deg\ to 80\deg\ show the presence of either \HII
regions or supernova remnants within the 2\deg\ $\times$ 2\deg\ region
centered at these positions, which suggests that the carbon
line emission might be associated with star forming regions.  Moreover,
these detections appear at velocities close to the tangent point
velocities at those longitudes.  The long path lengths near the tangent
points might have favored the detection of carbon lines in these
directions. Higher sensitivity observations of these regions should show
more detections in this longitude range if this is the case.  Indeed, our
high-resolution survey data has detected carbon lines at
several positions between $l = $ 20\deg\ to 38\deg\ as listed in
Table~\ref{tab:spectra}.  This clearly indicates that diffuse \CII regions
exist in this longitude range and the selection effects noted above are
likely responsible for their non-detections in the low-resolution survey.

\subsection{\lv diagram}
\label{sec:lvd}

\begin{figure*}
\sidecaption
 \includegraphics[height=10cm,width=12cm]{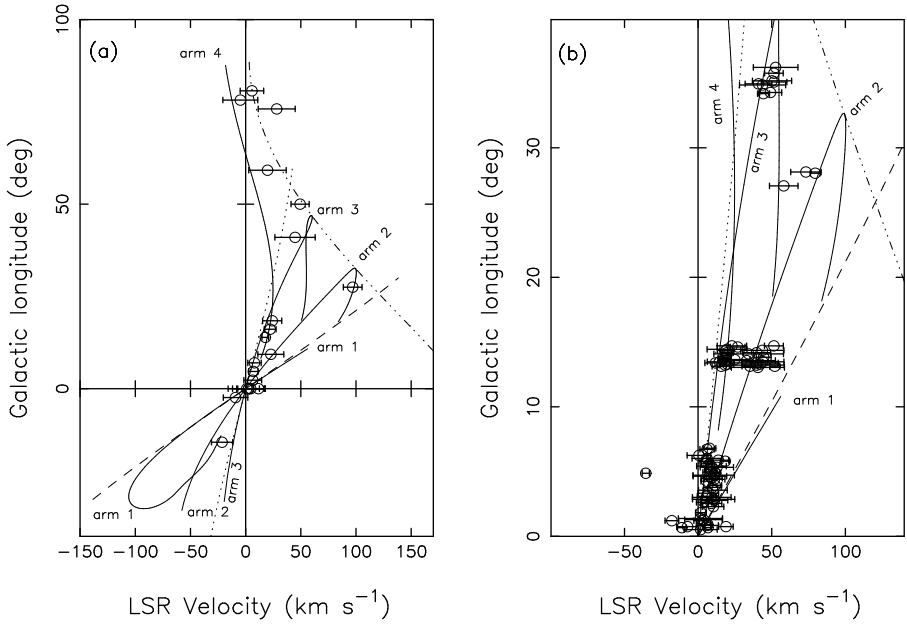}
 \caption{Longitude-velocity (\lv) diagrams constructed from carbon RRL
  emission at 327 MHz data: (a) using data from low-resolution (2\deg\ $\times$ 2\deg; paper I)
  survey; (b) using data from high-resolution (2\deg\ $\times$ 6\arcmin) survey.
  The marker indicates the central velocity whereas the length of the segment indicates
  the line width of the detected carbon lines.
  The four spiral arms (1 to 4 as designated by
  Taylor \& Cordes (\cite{tc93}) are shown as solid lines in each of the \lv
  diagrams. The dashed and dotted lines in each frame
  correspond to gas at galactocentric distances of 3.7 kpc and 7 kpc
  respectively.  The dash-dot-dot-dot-dash line indicates the locus of
  tangent points. }
  \label{fig:lvdiag}
\end{figure*}

\begin{figure}
 \resizebox{\hsize}{!}{\includegraphics{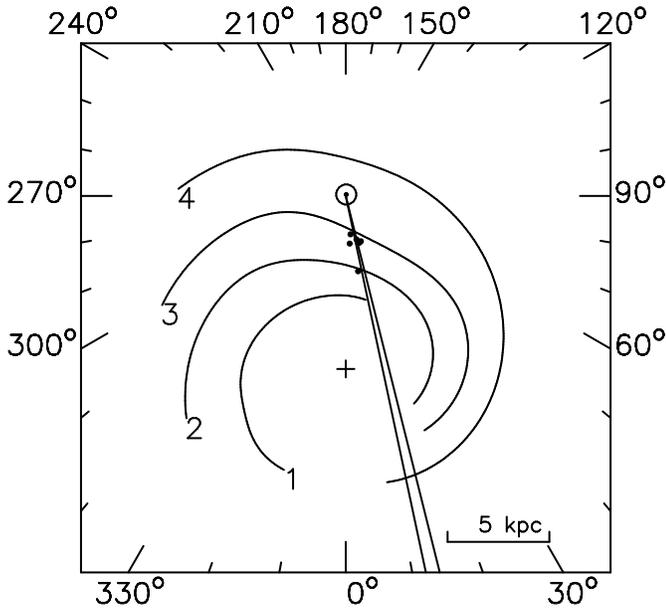}}
 \caption{Locations of the carbon line emitting regions
  (filled circles) between $l = $ 4\deg\ to 20\deg\ are shown in
  galactocentric coordinates. The regions are placed at the near kinematic distances
  estimated using the observed central velocity of carbon line emission.
  Most of the detected carbon line forming regions in this longitude range lie in spiral arm 3.
  The region of the galactic disk covered by the 2\deg\ wide field centered at G13.9+0.0 is also
  shown. This region intercepts spiral arms 3, 2 and 4 at a distance
  $\sim$ 1.9, 3.7 and 14.1 kpc from the Sun.}
  \label{fig:galdis}
\end{figure}

The longitude-velocity diagram constructed from RRL observations of
the galactic plane can be used to understand the distribution of the
carbon line-forming gas in the galactic disk if we make the standard assumption
that the observed central velocity of the line is due to differential
galactic rotation.  The \lv diagrams plotted for the
low-resolution and high-resolution survey data (Fig.~\ref{fig:lvdiag}a \& b)  show that
the carbon line emission arises from gas located at galactocentric distances beyond 3.7 kpc.
The line-forming gas at longitudes $\le$50\deg\ is confined 
between galactocentric distances of 3.7 kpc and 7.0 kpc.
Moreover, line emission in the low-resolution survey for longitudes $\le$50\deg\ shows, 
in general, some confinement to the spiral arms.  
The galactic rotation model used here has been taken from 
Burton \& Gordon (\cite{bg78}) after scaling it to \RO = 8.5 kpc and \thetaz = 220 \kms.  
Fig.~\ref{fig:galdis} shows the location of the line-forming regions obtained
from the low-resolution survey in the plane of our Galaxy between 
$l = $ 4\deg\ to 20\deg\.  These regions have been placed
at the near kinematic distance.  This is a reasonable assumption
since the large beam width (2\deg\ $\times$ 2\deg) of the low-resolution survey 
is likely to make the observations more sensitive to nearby regions.  
From the figure, it appears that most of the carbon line-forming gas in this longitude range
is associated with spiral arm 3.  Only toward   
$l = $ 9\deg.3, the near kinematic distance places the line emitting region near spiral arm 2. 
No line emission is detected from spiral arm 1 in this longitude range. 
In the high-resolution survey, line emission is detected
over a wider velocity range between $l = $0\deg\ and 40\deg\ 
compared to that in the low-resolution survey (Fig.~\ref{fig:lvdiag}). 
In general, the velocity range over which carbon lines near 327 MHz are detected in the 
surveys is similar to the velocity
spread of spiral arm tracers, for example, hydrogen RRLs near 3cm from \HII\ regions 
(Lockman \cite{l89}).  
No line emission is detected from 
spiral arm 4 in the longitude range 20\deg\ to 89\deg\ in both surveys. 
A few line detections in this longitude range have velocity close to the tangent points.
This is also a feature seen in the \lv diagram of spiral arm components in this longitude range
(see, for example, 3cm RRL emission from \HII regions; Lockman \cite{l89}).  
In summary, the \lv diagram of carbon line emission displays several similarities
with those of spiral arm tracers.

\clearpage

\clearpage

We compared 
the \lv diagrams obtained from the 327 MHz survey with those obtained from
the carbon absorption line data near 76 MHz (Erickson \etal \cite{ema95}) and 35 MHz 
(Kantharia \& Anantharamaiah \cite{ka01})
since the observations at these three frequencies overlap in the longitude range
$l =$ 332\deg\ $\rightarrow$ 20\deg.  The \lv diagrams show similar features. 
At all the three frequencies, most of the detections are at longitudes $< 20$\deg.
The \lv diagrams obtained from
the three observations indicate that the detected carbon line forming regions are confined 
between galactocentric distances of 3.7 to 8 kpc suggesting that they arise in
the same diffuse \CII regions. 
However, the width of lines detected
in absorption in many cases are larger (up to a factor of 2) than that of emission lines 
observed in the low-resolution survey.  The different line widths can be due
to (a) different beam widths of the surveys and (b) effect of pressure and radiation broadening
which have a strong dependence on the principal quantum 
number ($\alpha$ $n^{8.2}$ and $n^{8.8}$ respectively for widths in \kms; Shaver \cite{s75}). 
Interestingly, the width of the absorption line seems to extend over the velocity range 
over which emission lines are observed in the high-resolution survey at the corresponding 
longitudes.  Absorption lines near 76 MHz have been detected extensively at longitudes 
340\deg $ < l < 360$ \deg\
for which we have few detections near 327 MHz.  This is likely a case of
lack of sensitivity (see Section \ref{sec:lvsl} for more details) 
than any intrinsic property of the line-forming regions. 
The general similarity of the \lv diagrams obtained from the three observations indicates 
that the carbon lines observed near 76 MHz and 35 MHz are the
absorption counterparts of the carbon lines detected in emission near 327 MHz.

\subsection{Radial Distribution}
An \lv diagram gives a qualitative understanding of the
distribution of ionized gas in the galactic disk.  However, a more quantitative
study can be made by computing the average emission as a function of the
galactocentric radius.  Since the ionized gas at ``near'' and ``far''
kinematic distance will be at the same galactocentric distance, the
radial distribution is not affected by the two-fold ambiguity in
estimating the line-of-sight distance.  However the distribution will
depend on several other factors: (a) the sensitivity of the observations to
line-forming regions at different distances along the line-of-sight; 
(b) amplification of line intensity
due to stimulated emission by galactic non-thermal background; 
(c) choice of the rotation model used for the computation.

\begin{figure}
 \resizebox{\hsize}{!}{\includegraphics{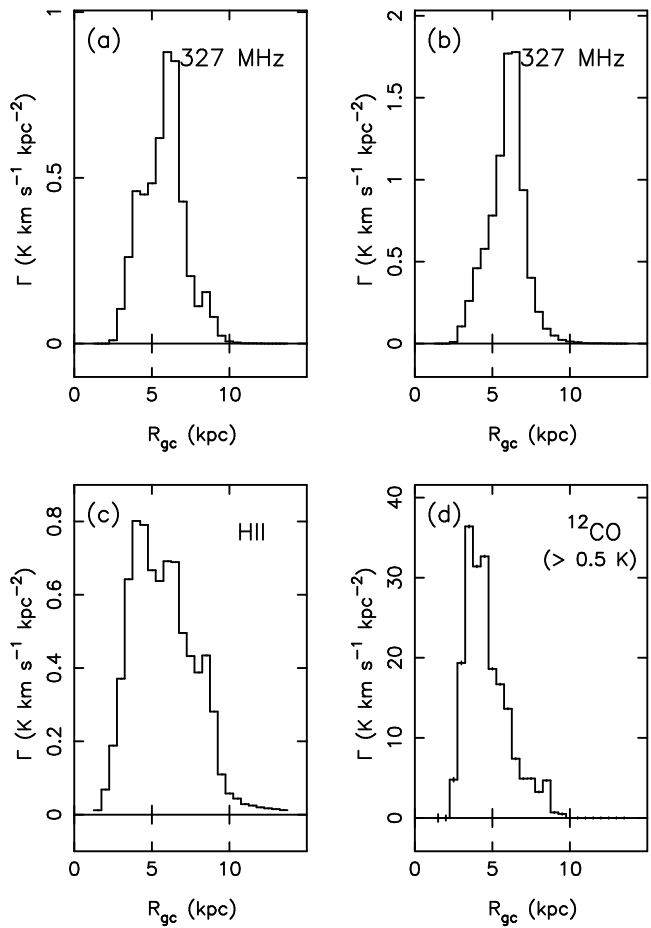}}
 \caption{The radial distribution
(average emission $\Gamma$ vs galactocentric radius \RGC)
of different components of the ISM is shown in the figures.
The radial distribution of (a) carbon RRL emission from the
galactic plane
near 327 MHz, (b) carbon RRL emission from the galactic plane
near 327 MHz in the longitude range $l = $ 4\deg\ to 20\deg,
(c) hydrogen  RRL emission from \HII regions near 3 cm and (d)
``intense'' (T$_A >$ 0.5 K) \CO emission from the galactic plane.  The
radial distributions in (a), (c) \& (d) were computed using the data in the longitude
range 4\deg~$< l <$ 84\deg\ where all the components of the ISM are
well sampled. The data are taken from Paper I (327 MHz carbon RRL), Lockman (1989)
(RRLs from \HII regions) and Dame \etal (\cite{detal87})(\CO).}
\label{fig:rdis}
\end{figure}

The radial distribution of the different traces of the interstellar medium (Fig.~\ref{fig:rdis}) 
are computed using the method described in Paper I.
In the computation for the carbon lines near 327 MHz,
the Gaussian fits to the observed profiles were used instead of the actual spectra. 
This was necessary since the typical peak line intensity to RMS noise for a carbon line detection is
only $\sim$ 3 to 4.  Using the Gaussian fit profile also
eliminates any contamination from the hydrogen line emission,
particularly for \RGC $<$ 2 kpc. 
We have used the carbon line data from the low-resolution survey 
between $l = $ 4\deg\ to 84\deg\ in the computation since
in this longitude range other components of the ISM (\HII regions
and \CO emission) are well sampled and hence a direct comparison 
of their distribution with the carbon line data is possible.

The radial distribution obtained from the low-resolution survey carbon line data (see
Fig.~\ref{fig:rdis}a) shows that the average emission extends from
\RGC = 2.5 kpc to 9 kpc with a prominent peak near 6 kpc. 
About 90 \% of the total observed carbon line emission originates  
between galactocentric distance 3.7 kpc and 8 kpc. 
The distribution falls off steeply on either side of the 6 kpc peak, the half width being 3.0
kpc.  However, the true distribution is likely to be narrower than this because the
broadening of the distribution due to intrinsic velocity dispersion has
not been taken into account.  
An increase in line emission near 8.5 kpc is also seen which is due to the Cygnus loop region 
in the nearby Orion arm.

The spiral arm structure in the galactic disk should be evident in the radial distribution
if the line emission shows some confinement to the spiral arms.
In Fig.~\ref{fig:rdis}(b), the carbon line distribution
computed using data in the longitude range, $l = $ 4\deg\ to 20\deg\ is shown.
Since most of the carbon line emission we detect is from this longitude range,
it resembles the distribution in  Fig.~\ref{fig:rdis}(a) with the prominent peak near 6 kpc   
clearly seen and the small peak near 8.5 kpc missing.  
As discussed in Section.~\ref{sec:lvd}, the line emission in this longitude range 
is likely to be confined to spiral arm 3, which naturally explains the peak at 6 kpc
since the average distance to the spiral arm is $\sim$ 6 kpc. Comparing 
Figs.~\ref{fig:rdis}(a) and (b), it is seen that (a) shows slight excess 
emission near 4 kpc.  Although there is no prominent peak in our low resolution data 
(Fig.~\ref{fig:rdis}(a)) there is some carbon RRL emission 
at 4 kpc distance associated with spiral arm 2.
Moreover, we do detect emission from spiral arm 2 from several positions within this 
longitude range in our high resolution survey data (see Fig~\ref{fig:lvdiag}b).
Future higher resolution, sensitive observations are required to check the widespread presence
of carbon line emission in spiral arm 2 in the inner Galaxy.

We compared the radial distribution of the carbon line emission with other components
of the ISM to check for any similarities that may exist.  We find that
the radial distribution of carbon lines is distinct from that
of \HI. 
The latter is observed up to the outer reaches of the Galaxy (Burton \cite{b88})
whereas the carbon line emission is confined to galactocentric distances between
2.5 kpc to 9 kpc with well-defined peaks in its radial distribution. 
Comparing the radial distribution of carbon line emission 
with the distribution of the 3 cm hydrogen RRL emission from compact
\HII regions and ``intense'' \CO emission  (Fig.~\ref{fig:rdis}c \& d),
both spiral arm tracers (Solomon, Sanders \& Rivolo \cite{ssr85}), 
we find a number of similarities.  
Both, the 3 cm hydrogen RRL emission and \CO emission are confined
(see Fig.~\ref{fig:rdis}c \& d; for details see Paper I) 
in the range \RGC = 2.5 kpc to 9 kpc which is similar to the 
carbon line emission.  A peak near 6 kpc is seen in the
distribution of 3 cm hydrogen RRL emission and considerable \CO\ emission
is present at the radial distance of 6 kpc, which
is similar to that seen in the distribution of carbon line emission. 
We conclude that the carbon line emission near 327 MHz 
has similar galactic disk distribution as that of the star-forming regions.  
This result may appear somewhat different 
from what we know about the gas toward CasA $-$ where the morphology
of the carbon line forming gas resembles the distribution of \HI observed in absorption across Cas A
(Anantharamaiah \etal \cite{aepk94}) and no hydrogen RRL 
has been detected (Sorochenko \& Smirnov \cite{ss93}).  However, it is not contradictory since 
in the inner Galaxy, the distribution of \HI observed in absorption 
resembles that of \CO.  \HI with $\tau_{HI} > $ 0.1 
shows an 85 \% probability of being associated with \CO emission 
(Garwood \& Dickey \cite{gd89}).

\begin{figure}[b]
\includegraphics[height=11cm]{MS2320f6.ps}
\includegraphics[height=3cm]{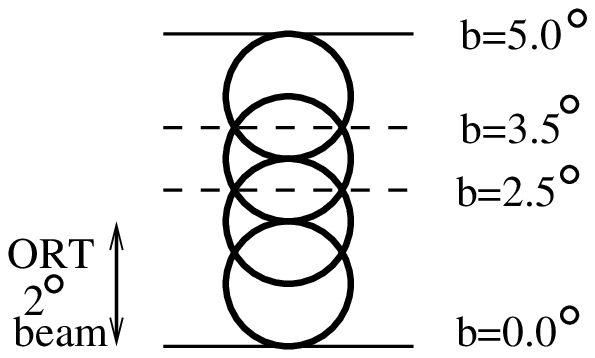}
 \caption{
Spectra of data averaged over different galactic latitude ranges toward
$l = $ 0\deg.0 are shown in the top plot.  The bottom schematic shows the 
stacking of the ORT beams along positive latitudes toward 
$l = $ 0\deg.0; similar data was also obtained
for negative galactic latitudes.  The spectra in the panels:
a) G0.0+3.5avg was obtained by averaging data from 
beams centered at $b =$ 3\deg, 4\deg $~~$  b) G0.0+2.5avg was obtained by averaging data over 5\deg$~$
from beams centered at $b =$ 1\deg, 2\deg, 3\deg, 4\deg. $~$ c) G0.0+0.0avg was 
obtained by averaging data over 10\deg\ centered
at $b =$ 0\deg.  d) and e) are similar to b) and a) except that the data is averaged over
negative latitudes.  The spectrum toward the galactic centered has been
excluded from all the averaged spectra. The carbon line emission is seen to extend from $b\sim -$2\deg
to $\sim$4\deg. }
\label{fig:lat1}
\end{figure}

\begin{figure}
\includegraphics[height=14cm]{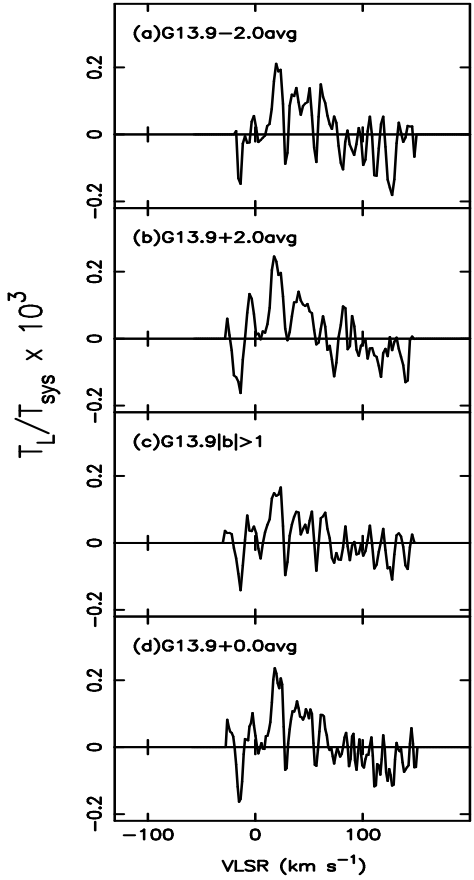}
\caption{
Spectra obtained by averaging data over different galactic latitude range toward $l = $ 13\deg.9.
Data is obtained up to $b=\pm3$ \deg.   
a) G13.9-2.0avg is obtained by averaging the data over $4$ \deg in latitude about $b = -2$\deg. 
b) G13.9+2.0avg is same as (a) but about $b = +2$\deg.
c) G13.9$|b|>$1 is the spectrum obtained by averaging all the data
at latitude $|b| >$ 1\deg~. 
The spectrum toward $l=$13\deg.9, $b=$0\deg.0 has been excluded
in all the averaged spectra shown in the figure. }
\label{fig:lat2}
\end{figure}

\section{\FIRCII line emission from carbon RRL forming region}
\label{finestructure}

The \FIRCII line is due to the radiative decay of the fine structure
transition $2P_{3/2} \rightarrow 2P_{1/2}$ in singly-ionized carbon. 
Recombination lines of carbon are a result of electronic transitions in a
recombined atom in ionized gas.  The excitation temperature 
of the fine structure transition
($\sim 91$ K) is comparable to a subset of the temperatures which explain the
observed low frequency carbon RRL emission (Kantharia \& Anantharamaiah \cite{ka01}). 
Moreover, dielectronic-like recombination (Watson, Western \& Christensen \cite{wwc80}) 
is a process
involving the excitation of the fine-structure levels  which modifies the electronic 
level populations in recombined carbon; thus modifying the observed line optical
depths of the carbon recombination lines.
Since the two emission mechanisms are intricately linked,
it is interesting to study their correlation.  

In this section,
we estimate the expected \FIRCII emission strength from low frequency
carbon RRL-forming regions and compare the 
galactic distribution of the diffuse \FIRCII fine structure line
with the carbon RRLs detected near 327 MHz. 

\subsection{The \FIRCII emission from the 327 MHz carbon RRL forming regions}

The \FIRCII line originates predominantly from three types of regions:
photodissociation regions (PDRs), cold neutral medium (CNM) and extended low-density 
warm ionized medium (ELDWIM) (Petuchowski \& Bennett \cite{pb93}, 
Heiles \cite{hei94}).  
As described by Hollenbach \etal (\cite{htt91}), carbon is mostly
in singly-ionized state upto $A_V <$ 4 mag in 
low-density PDRs (the dense PDRs have a
relatively low volume filling factor and hence may not contribute largely
to the global diffuse \FIRCII emission).  Hence low-density PDRs,
which for the present discussion are considered as regions associated with 
molecular clouds, 
are possible sources of the diffuse fine structure line emission as well as  
carbon RRL emission.
The CNM is another source of singly-ionized carbon. 
The CNM is distinct from PDRs in that they are predominantly atomic clouds
with 
neutral densities $< 10^3$ \cmthree and 
typically $A_V \le$ 1 mag (Heiles \cite{hei94}). 
The \FIRCII line is the major cooling
transition in the CNM and the low-density PDRs since the number density of
the colliding particles is generally less than the critical density, which
depends on the colliding particles and their temperature. 
For temperatures relevant for CNM and low-density PDRs ($\sim$ 20 to 500 K), 
the critical densities due to collision with atoms and molecules are $\sim$ 3000 \cmthree 
and $\sim$ 4000 \cmthree respectively
(Launay \& Roueff \cite{lr77}, Flower \& Launay \cite{fl77}). 
For densities larger than these, the fine-structure level is collisionally de-excited. 
In the ELDWIM, which consists of both the warm ionized medium (WIM;
Reynolds \cite{r93}) and 
low-density ($n_e \sim$ 1 -- 10 \cmthree) ionized gas in the inner Galaxy 
(Petuchowski \& Bennett \cite{pb93},
Heiles \cite{hei94}), carbon is expected to be ionized.
The critical
density for collisions with electrons is $\sim$ 30 \cmthree (Hayes \& Nussbaumer \cite{hn84}) 
assuming a temperature of 7000 K for the low-density ionized component (Anantharamaiah \cite{a85}). 
Roshi and Anantharamaiah (\cite{ra01b}) calculated
a contribution of 8.1$\times 10^{-5}$ 
ergs s$^{-1}$ cm$^{-2}$ sr$^{-1}$ from the low-density ionized 
regions (one of the components of ELDWIM) 
in the longitude range $l = $ 0\deg\ to 20\deg\ (relevant for
the comparison between carbon RL and far-infrared line emission).  The
{\it diffuse} \FIRCII emission within $|b| <$ 2\deg\ obtained from
the higher resolution far-infrared line observations is
$\sim$ 1.5 $\times 10^{-4}$ ergs s$^{-1}$ cm$^{-2}$ sr$^{-1}$
(Nakagawa \etal \cite{naka98}).   Thus, the ELDWIM can contribute
$\ge 54\% $ of the observed \FIRCII emission between 
$l = $ 0\deg and 20\deg.
However, ELDWIM is not a dominant source of 327 MHz carbon RRL emission since 
its temperature is high (line optical depth
$\propto T_e^{-2.5}$) and carbon abundance is only depletion factor
times the cosmic abundance ($4 \times 10^{-4}$; Spitzer \cite{s78}). 
Moreover, the ratio of the carbon to hydrogen line intensity
detected in the 327 MHz survey is $\sim$ 0.5 which is much higher than
what is expected from the abundance ratio, suggesting a distinct origin
for the two lines. The CNM and PDRs with their relatively low temperatures 
are envisaged as likely sites of origin for the 327 MHz carbon RRLs.  
Hence we estimate the contribution
of the carbon RRL forming CNM and PDRs to the observed \FIRCII line intensity.

For estimating the intensity of the FIR line from carbon RRL forming regions, 
we considered typical parameters
estimated for diffuse \CII regions.  Kantharia \& Anantharamaiah (\cite{ka01}) have 
modeled the diffuse \CII regions in a few directions in the inner Galaxy.  
They find that models with temperatures in the range 20 $\rightarrow$ 80 K
can fit the observations depending on the angular extent of the line
forming region. Even higher temperature ($\sim$ 150 K) models could fit
the observations. Since the total \HI column density (hence \HI opacity)
predicted by the higher temperature ($\sim$ 150 K) models are larger than 
that observed in the inner Galaxy, 
we use only models with temperatures in the range 20 $\rightarrow$ 80 K  
for the FIR line intensity calculation.
The estimated electron density and path length corresponding to the 
observed integrated optical depth near 327 MHz of 
$\sim 0.01$ \kms~ in the inner Galaxy 
for this temperature range are $0.1 \rightarrow 0.03$ \cmthree and
$0.2 \rightarrow 20$ pc respectively.
The temperatures and electron 
densities which explain the low frequency carbon RRLs are
encountered in the CNM (Heiles \cite{h01}) as well as low-density PDR 
(Hollenbach \etal \cite{htt91}). If the line
emission is associated with the CNM then the neutral density is $\sim$ 
500 $\rightarrow$ 150 \cmthree, which is the atomic density in these clouds. 
We assumed a carbon depletion factor of 0.5 for
these estimates and other calculations presented here. The thermal 
pressure of these regions are $10000 \rightarrow 12000$
\cmthree K, which are not unreasonable for the CNM (Jenkins, Jura \& Lowenstein \cite{jjl83}).
The above numbers translate to hydrogen column densities ranging from 
$\sim 3.1 \times 10^{20} \rightarrow
9.3 \times 10^{21}$ cm$^{-2}$. 
Such column densities are not unreasonable in the inner Galaxy (Dickey \& Lockman \cite{dl90}).  However 
toward the higher end, they
cannot be reconciled with the observed width of the carbon lines
(since they have to be shared by different CNM clouds). We
discuss these issues in a later publication.
Here, we consider the above possible physical conditions for diffuse \CII regions 
coexisting with CNM. 

As described above, the physical properties of carbon RRL forming region is
also encountered in low-density PDRs. The regions with $A_V <$ 3 mag of low-density PDR models of
Hollenbach \etal (\cite{htt91}) have temperature similar to the higher temperature ($\sim$ 80 K)
models of carbon RRL forming regions. Hydrogen is mostly atomic in these regions of the PDR. 
The neutral density of these regions should be
$>$ 150 \cmthree to produce the required electron density ($>$ 0.03 \cmthree) needed for the 
carbon RRL forming region. 
Typical observed \HI column density of such regions associated with molecular
clouds is $\sim 10^{20}$  cm$^{-2}$ 
(Wannier, Lichten \& Morris \cite{wlm83}), which 
means several such low-density PDRs are needed along a sight-line to produce
the observed carbon RRL.  The low-temperature ($\sim$ 20 K) ``diffuse'' \CII regions
could be zones with $A_V \sim$ 4 mag of the PDR. For example, a low-density PDR
model with $n_0 \sim 10^3$ \cmthree and incident FUV flux of $\sim$ 1.6 ergs cm$^{-2}$ s$^{-1}$
can have gas temperature $\sim$ 20 K and electron density $\sim$ 0.1 \cmthree at 
$A_V \sim$ 4 mag (Hollenbach \etal \cite{htt91}). 
Hydrogen is mostly molecular in these regions of the PDR. 
For the estimation of FIR line emission from $A_V \sim 4$ mag
region, we use the above given parameters for the low-density PDR, which
are typical values in the inner Galaxy. The \HI density for this model is $\sim$
10 \cmthree and molecular density  is $\sim$ 1000 \cmthree (see Fig 4a of Hollenbach \etal \cite{htt91}).

The intensity of the \FIRCII line from the neutral regions
is given by (Bennett \etal \cite{ben94}, Watson \cite{wat84})
\be
I_{\CII} = 7.416\times 10^{-3}\frac{\frac{g_u}{g_l} e^{-h\nu/kT}}{1+\frac{g_u}{g_l} e^{-h\nu/kT}+ \left[ \Sigma \frac{n_i}{ncr_i} \right] ^{-1}} n_{C^+}L,
\ee
where $g_u (=4), g_l (=2)$ are the statistical weights of $2P_{3/2}$ and $2P_{1/2}$ states
respectively, $h\nu = 1.26 \times 10^{-14}$ ergs is the energy of the 158 $\mu$m photon, 
$k$ is the Boltzmann
constant, $T$ is the gas temperature in K, $n_{C^+}L$ is the column density of 
ionized carbon in \cmthree pc.   The above equation has been derived assuming a 
optically thin line from a two energy state atom.
The population of energy states are determined by collisions and spontaneous
emission in the optically thin case. In the above equation $n_i$ is the density
of colliding particles. $ncr_i$ is the critical density, which is defined
as the ratio of the collision rate to the spontaneous 
emission rate. $ncr_i$ depends on the temperature of the interacting particles. 
For the temperatures encountered in the carbon RRL forming region $ncr_i \sim$ 10 \cmthree
for electron collision (Hayes \& Nussbaumer \cite{hn84}) and 3000
and 4000 \cmthree for neutral hydrogen and molecular hydrogen
respectively as described earlier. 

We calculate the expected intensity of the \FIRCII emission from the CNM and PDR
at temperatures of 20 K to be 
$5.2 \times 10^{-7}$ ergs s$^{-2}$ cm$^{-2}$ sr$^{-1}$
and $6.2 \times 10^{-7}$ ergs s$^{-2}$ cm$^{-2}$ sr$^{-1}$ respectively.  
For temperatures of 80 K, the expected intensity of the fine-structure line from
CNM and PDR is found to be
$1.4 \times 10^{-4}$ ergs s$^{-2}$ cm$^{-2}$ sr$^{-1}$.
Comparing these estimates with what the Balloon-borne Infrared Carbon Explorer (BICE)
observed in the inner Galaxy for
$|b| < 1$\deg\ (Nakagawa \etal \cite{naka98}), it appears that for temperatures near 20 K, the
contribution to the total observed \FIRCII intensity is a negligible 0.4 \% whereas
if the temperatures of the \CII regions are near 80 K, then 95 \%
of the total observed \FIRCII intensity can arise in the diffuse \CII
regions coexistent with CNM or low-density PDRs.  
Thus, if the temperature of the carbon RRL forming regions is low
($\sim 20$ K), then most of the fine-structure line emission is likely
to arise elsewhere -- either in the ELDWIM or CNM and low-density PDR that
do not produce observable carbon RRLs.  If the temperature is
high ($\sim 80$K), then most of the fine-structure emission is likely
to come from the PDRs and CNM that form the same family of diffuse
\CII regions which give rise to the low frequency carbon RRLs. In that
case a more accurate estimate of the physical properties of the carbon
RRL forming region is required to determine the relative importance of
ELDWIM and PDRs/CNM to the global contribution of \FIRCII 
line emission. This will be attempted in future with multi-frequency
carbon RRL data.

We note that in the inner Galaxy the assumption that the \FIRCII emission
is optically thin is not entirely true.  The opacity of the
\FIRCII line is  $\sim 0.9$ for a typical carbon RRL width of 
14 \kms\ (Heiles \cite{hei94}) arising in a cloud with temperature 80 K in the inner Galaxy.  
However, for simplicity and to get a first order estimate, we have
considered the optically thin case which gives us the interesting results
discussed above.  

\subsection{Longitudinal distribution of the carbon FIR line and radio line}

We also attempted a comparison of the
longitudinal distribution of the two tracers of ionized carbon.  This
is relevant since, as discussed in the previous subsection,
a considerable fraction of the fine-structure
line can be accounted for by the diffuse \CII regions observed in low frequency
carbon RRLs under certain physical conditions.  
If the longitudinal distributions of the two tracers are similar,
it would support the higher temperature ($\sim 80$ K) models for
the carbon RRL forming regions and a substantial fraction of
the observed \FIRCII emission is likely to arise in the carbon RRL forming region.

Wright \etal (\cite{wri91}) and Bennett \etal (\cite{ben94}) have presented the galactic
distribution of the \FIRCII line with an angular resolution of
$\sim 7$\deg~using the data from the Far-Infrared Absolute Spectrophotometer
(FIRAS) aboard the Cosmic Microwave Background Explorer (COBE). 
Bennett \etal (\cite{ben94}) report strong \FIRCII emission in the galactic plane
with a half-intensity longitude range of $\sim$ 360\deg $\rightarrow$ 40\deg.
A peak in the FIR emission is seen near $l=80$\deg~ which matches with
a peak seen in our 327 MHz RRL data.  
Bennett \etal (\cite{ben94}) caution against over-interpreting 
this peak due to few measurements in that region.  
Although the angular resolutions of the two datasets are different,
a comparison of the gross distribution shows that the \FIRCII emission
is more widespread than the carbon RRLs near 327 MHz (see Fig \ref{fig:cltctl}).  
Nakagawa \etal (\cite{naka98})
have used the data from BICE with a much
finer angular resolution of $15'$.  
Their survey  
covers the region from $l = $
$350^{\circ} \rightarrow 25^{\circ}$.
They detect \FIRCII emission in this longitude range from both ``compact'' and ``diffuse'' regions.
The ``diffuse'' \FIRCII emission is observed to extend almost uniformly till the longitude
limits of their observations in the galactic plane.  
In slight contrast, carbon RRLs near 327 MHz have been detected almost contiguously 
between $l = $ 0\deg to 20\deg~.   
We do not detect carbon RRLs in the fourth quadrant 
(up to $l \sim -15$\deg) due to reduced
sensitivity of the equatorially-mounted Ooty Radio Telescope. 
Nakagawa \etal (\cite{naka98}) also observed reduced \FIRCII emission in regions adjacent to the
galactic center up to about longitudes $\sim \pm 4$\deg~.  This is a 
behavior distinct from our low-resolution survey carbon RRL 
data which shows comparable integrated
optical depths in the galactic plane from $l=0$\deg~ till $l=+10$\deg~ 
(see Fig \ref{fig:cltctl}).  

In summary, intense FIR emission in the galactic plane is observed 
in the longitude range where carbon
RRL near 327 MHz is detected. But the FIR emission seems to be more widespread
in the galactic plane. Note that the comparison is, however, limited by (1) the
large difference in the sensitivity of the FIR and carbon RRL observations; (2)
poor velocity resolution of the existing FIR data. A comparison of the LSR velocities
of the two tracers is essential in further establishing any connection
between the two spectral lines. We therefore conclude that the existing data
does not rule out the possibility that the ``diffuse'' \CII regions can significantly
contribute to the \FIRCII 158 $\mu$m emission in the inner region of the Galaxy. 
   
\section{Latitude extent of Carbon line emission}
\label{sec:lat}

We examined the data collected along galactic latitude toward two longitudes
($l = $ 0\deg.0, Fig~\ref{fig:lat1} and $l = $ 13\deg.9, Fig~\ref{fig:lat2}) in the low-resolution survey.  
The beam center was shifted by half the beamwidth (\ie 1\deg) along the galactic
latitude and data was collected up to $b = \pm$ 4\deg\ toward $l=0$ \deg\ and
up to  $b = \pm$ 3\deg\ toward $l= $13\deg.9. 
The carbon feature is clearly seen in the spectrum averaged over 
the entire latitude range observed toward $l=0$\deg\ (see Fig.~\ref{fig:lat1}c).  
The data toward the Galactic center was excluded since the carbon
line emission in this direction is fairly strong and hence likely to
dominate the averaged spectrum.
Carbon lines are also detected when the data from positions separated by the beamwidth 
(\ie 2\deg) over the entire sampled region are averaged, confirming 
the presence of carbon line emission over several degrees in 
galactic latitude.
\clearpage
\begin{table}
\caption[]{Summary of the study of latitude extent of carbon line emission 
\label{tab:lat}}
\begin{tabular}{lrrrccc}
\hline  \hline
\multicolumn{1}{c}{$Position$} & \multicolumn{1}{c}{$T_L/T_{sys}$\footnotemark[1]} & \multicolumn{1}{c}{$\Delta  V$} & \multicolumn{1}{c}{$V_{\rm{LSR}}$} &  $V_{res}$\footnotemark[4] & RMS\footnotemark[2] &\multicolumn{1}{c}{$t_{int}$}  \\
       &\multicolumn{1}{c}{$\times$ 10$^{3}$}&\multicolumn{1}{c}{\kms} &\multicolumn{1}{c}{\kms}     & \kms     & $\times$ 10$^{3}$ & \multicolumn{1}{c}{hrs}  \\ 
\hline
\multicolumn{7}{c}{\bf Observations towards $l = $ 0\deg.0} \\
\hline
    G0.0$-$2.5avg & 0.13(0.02) & 39.3(6.1) &    4.3(2.6) & 4.8 & 0.03 & 44.6   \\
    G0.0$+$2.5avg & 0.32(0.03) & 15.6(1.8) &    2.9(0.7) & 3.4 & 0.05 & 49.6   \\
    G0.0$-$3.0avg & & & & 7.6 & 0.03 & 32.1  \\
    G0.0$+$3.0avg & 0.27(0.04) & 12.5(2.2) &    3.3(0.9) & 3.4 & 0.06 & 37.7   \\
    G0.0$-$3.5avg & & & & 4.8 & 0.07 & 19.7  \\
    G0.0$+$3.5avg & 0.34(0.06)\footnotemark[3] & 10.3(2.0) &    2.8(0.8) & 3.4 & 0.07 & 26.1   \\
    G0.0$+$0.0avg & 0.20(0.02) & 23.6(2.5) &    2.7(1.0) & 3.4 & 0.03 & 94.2   \\
\hline
\multicolumn{7}{c}{\bf Observations towards $l = $ 13\deg.9} \\
\hline
   G13.9$-$2.0avg & 0.27(0.09)\footnotemark[3] & 9.2(3.5) & 19.8(1.5) & 4.8 & 0.09 & 29.0  \\
   G13.9$+$2.0avg & 0.25(0.05) & 11.0(2.5) &   18.5(1.1) & 4.8 & 0.05 & 30.8   \\
   G13.9$|b|>$1 & 0.15(0.05)\footnotemark[3] & 10.0(3.8) &   18.5(1.6) & 4.8 & 0.05 & 38.6   \\
   G13.9$+$0.0avg & 0.23(0.04) &  9.5(2.0) &   18.9(0.8) & 3.4 & 0.05 & 59.8   \\
               & 0.12(0.03) & 26.6(6.9) &   44.6(2.9) & 3.4 & 0.05 & 59.8   \\
\hline
\end{tabular}
\end{table} 
\footnotetext[1]{The line intensities are given in units of $T_L/T_{sys}$,
where $T_L$ is the line antenna temperature and $T_{sys}$ is the
system temperature}
\footnotetext[2]{RMS is in units of $T_L/T_{sys}$.}
\footnotetext[3]{tentative detection.}
\footnotetext[4]{ The resolution to which the spectra are
smoothed for estimating line parameters. }

\clearpage
Our data indicates that the carbon line emission extends 
from $b \sim -$2\deg\ to $b \sim +$4\deg\ toward the galactic longitude $l = $0\deg.  
The widths of the carbon lines seen in the positive latitude spectra and
the negative latitude spectra (Fig.~\ref{fig:lat1}b \& d) differ
by a factor of $\sim 2.5$.  The line parameters are listed in Table~\ref{tab:lat}.
The difference in line width
may indicate the presence of distinct line emitting regions along the latitude extent, 
maybe with different physical properties.  

Fig.~\ref{fig:lat2}d shows the spectrum toward the longitude $l = $13\deg.9 
obtained by averaging the line emission along latitude between $b = \pm$ 3\deg\ .
The spectrum toward $l=$13\deg.9, $b=0$\deg\ has been excluded in the average
spectra shown in Fig.~\ref{fig:lat2}.
A narrow feature ($\Delta V \sim $ 9 \kms) is clearly detected near  18.5 \kms.
A weak, broad carbon line feature also seems to be present in the spectrum.
The spectra averaged over the positive and negative latitude extents are shown in
Figs.~\ref{fig:lat2}a \& b.  
The narrow feature is clearly evident in these spectra; however
the signal-to-noise ratio of this feature is low. 
This feature is also present in the spectrum shown in Fig.~\ref{fig:lat2}c,
which is the average of all data at $|b| > $1\deg. 
The line parameters are listed in Table~\ref{tab:lat}.
The narrow carbon line emission is extended over, at least, 
the latitude range $-3$\deg\ to $+3$\deg\ suggesting the 
presence of a single large diffuse \CII\ region.
(see Section~\ref{sec:angG13} for further discussion 
on the narrow carbon line emission.)

From the above two cases, we believe 
that the carbon line emitting gas in the inner Galaxy is spread 
over a galactic latitude extent of at least $b \sim \pm3$\deg.  

\section{Angular extent of carbon line-forming regions}
\label{sec:ang}
Since only coarse resolutions are
available at the low frequencies that diffuse \CII regions can 
be studied using recombination
lines and observations at different frequencies have different angular
resolutions, it has been difficult
to obtain definitive estimates of the angular and linear
sizes of the diffuse \CII regions.  The ambiguity in the distances to 
these regions and the uncertainty in the angular size  makes it difficult
to obtain strict constraints on the linear size.  Kantharia \& Anantharamaiah (\cite{ka01})
modeled the carbon line data at three frequencies (35, 76 and 327 MHz) and obtained
different physical models for different angular sizes of the line-forming region.
They also attempted interferometric imaging of one position in the galactic plane
in carbon recombination line using the VLA to obtain the angular extent of the line
forming region.  They obtained a lower limit on the angular size of $10'$. 
Clearly the angular size is an important parameter entering into the modeling of
these regions and needs to be understood better. 
Since carbon lines are detected extensively in our low-resolution survey, it is
likely that the diffuse \CII regions are either 2\deg\ or more in angular extent or else consist
of several small $ \le 2$\deg\ clumps within the beam.  
In this section we try to answer the question ``do the
line-forming regions consist of clumps with emission
confined to small angular regions or is the emission extended 
and uniform over a large area ? '' We make use of the high-resolution
data to answer this question.  

\begin{figure}
\includegraphics[height=10cm]{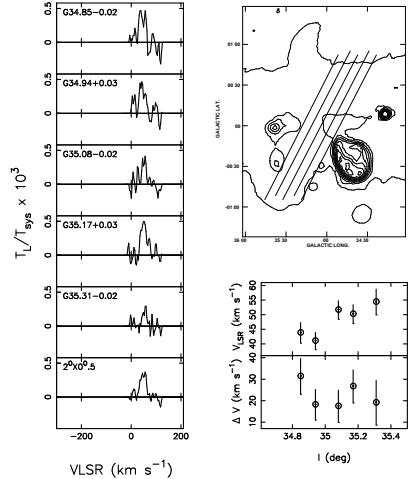}
\caption{Carbon line emission near $l = $ 35\deg.  Top five
panels on the left show the spectra observed with an angular resolution
of 2\deg\ $\times$ 6\arcmin\ toward the positions marked on each frame and
the lowermost panel shows the spectrum obtained by
averaging the five spectra. The 
observed positions are marked by the slanted lines on the 11 cm continuum
map from Reich \etal (\cite{rfrr90}).  The slanted lines represent the 2\deg $\times$ 6\arcmin
ORT beam.  The LSR velocity and the width
of the lines as a function of galactic longitude are shown
in the plot on the right-hand side. The vertical bar represents the 3 $\sigma$ 
error in the estimated parameters.
 }
\label{fig:aextenta}
\end{figure}

\subsection{Clumps in the diffuse \CII regions}

If the line emission arises in a homogeneous region with 
an angular extent of several degrees, then the line parameters observed
at positions within this angular span are expected to be
similar.  Examination of the observed spectra in the high-resolution survey
shows that at several positions there is considerable change in the line 
parameters when the beam center is shifted by $\sim 6'$ in declination.
For example, the width of the observed carbon line toward the position G5.19+0.02
is $\sim 8$ \kms\ is about one-third the line width observed toward
G5.33$-$0.03 ($\sim $ 23 \kms).  The beam centers of the two positions 
are separated by $\sim 9$\arcmin.  
Another example  is toward the direction $l =$ 35\deg.1 and 
$b = $ 0\deg (see Fig.~\ref{fig:aextenta}).
Carbon line is clearly detected in the integrated spectrum obtained by averaging 
the high-resolution survey data over the longitude range 
$l = $ 34\deg.85 to 35\deg.31 (0\deg.5 (along $l$) $\times$ 2\deg\ (along $b$) region). 
However on  examining the five  
contiguous spectra (observed with a 2\deg $\times$ 6\arcmin\ beam)
separated by $\sim 6$\arcmin, 
we find that 
the lines at positions with $l < $ 35\deg\ have different central velocities 
compared to those at positions with $l >$ 35\deg\ (see Fig.~\ref{fig:aextenta}).
Such behavior is exhibited toward
many other positions separated by $\sim 6$\arcmin. 
This suggests that line emission in these directions arises from 
distinct diffuse \CII regions or else that the diffuse \CII regions
have sub-structure on scales of  $\sim$ 6\arcmin. 
The near kinematic distance 
corresponding to the central velocity (48.6 \kms) of the integrated spectrum 
(Fig.~\ref{fig:aextenta}) is 3.3 kpc, 
which is close to the line-of-sight distance to the spiral arm 3 at this longitude. 
If the angular extent of the clump at $l <$ 35\deg is 
$\sim$ 6\arcmin\ 
then it
corresponds to a linear size $\sim$ 6 pc at the near kinematic distance. 
It, therefore, is likely that the diffuse \CII regions toward G35.1+0.0 consist of such
small line-forming clumps.

\subsection{Extended diffuse \CII regions}
A subset of our data also shows a behavior different from what we discussed in the previous section.  
The high resolution data within
the longitude range  $l = $ 1\deg.75 to 6\deg.75 (within Field 2) seems to indicate the occurrence of
a single \CII region extended over a region of angular size $\sim 5$\deg\
in longitude.  A similar extended ($\sim$ 2\deg $\times$ 6\deg) \CII\ region
is also observed toward $l = $ 13\deg.9, $b = $0\deg\ (Field 3). These 
extended \CII\ regions are discussed in detail below.

\subsubsection{Carbon Line Emission toward Field 2}
\label{sec:angG13}

\begin{figure}
 \resizebox{\hsize}{!}{\includegraphics{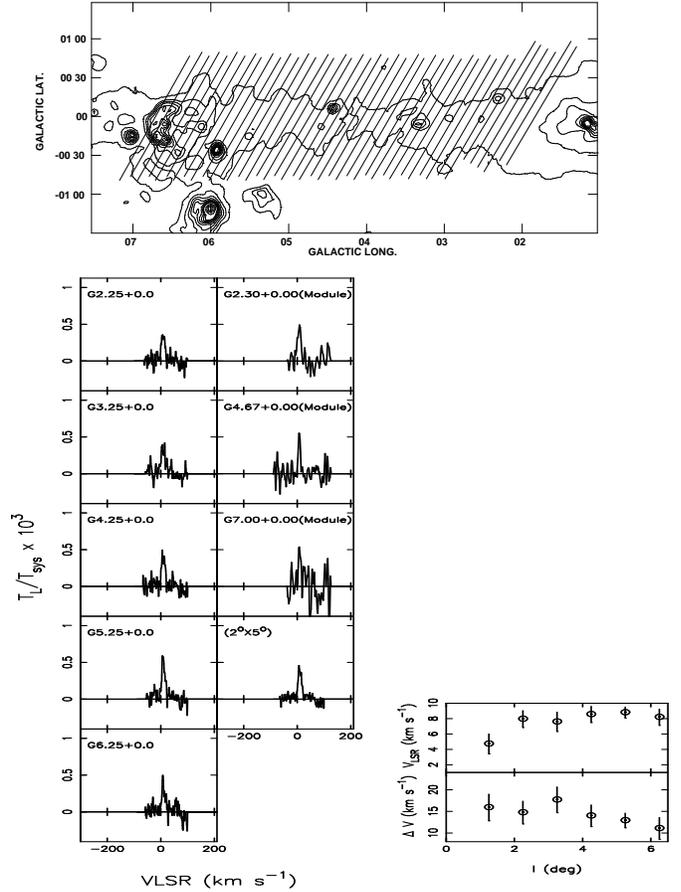}}
\caption{
Carbon line emission toward a 5\deg\ (along $l$) wide region 
centered at $l = $ 4\deg.25 and $b = $ 0\deg.
The spectra shown in the panels in the first column are obtained by averaging the 
high-resolution (2\deg $\times$ 6\arcmin) survey data
over a 2\deg $\times$ 1\deg\ area centered on the galactic 
coordinates indicated in each frame. The top three spectra
in the second column are from the low-resolution 
(2\deg $\times$ 2\deg) survey, observed toward the galactic coordinates
indicated in each frame. The lowermost spectrum in the second column is obtained
by averaging the high-resolution survey data over the 5\deg\ wide region. 
The LSR velocity and the width
of the lines from the spectra averaged over a 2\deg $\times$ 1\deg\ region,
as a function of galactic longitude are shown
in the right-hand side plot. The vertical bars represent the 3 $\sigma$ 
error in the estimated parameters.
The narrow feature near 8 \kms\
is observed in all the spectra indicating that the 
line-forming region is fairly extended in the sky plane. The 
observed positions are marked by the slanted lines on the 11 cm continuum
map from Reich \etal (\cite{rfrr90}). The slanted lines represent the 2\deg $\times$ 6\arcmin
beam of the ORT. 
}
\label{fig:aextentc}
\end{figure}

Here we examine the line emission seen from part of the 6\deg\ wide 
Field 2 ($l = $ 1\deg.75 to 6\deg.75) that we mapped using the high resolution data.
Most of the high-resolution spectra from this region detected a
$\sim$ 14 \kms\ wide carbon line centered on $\sim 8$ \kms.  To improve the signal-to-noise ratio 
of the line emission from the extended \CII\ region, we averaged the spectra over a region  
1\deg\ (along $l$) $\times$ 2\deg\ (along $b$). 
The averaged spectra are shown in Fig.~\ref{fig:aextentc}.  
The $\sim$ 14 \kms\  wide component is clearly seen in all the spectra.
Table~\ref{tab:lextentb} gives the line parameters obtained from Gaussian fits 
to the spectra.  The large angular extent of the $\sim$ 14 \kms\ wide component
is also evident from the
detection of this component in the low-resolution observations toward G2.3+0.0,
G4.7+0.0 and G7.0+0.0 (see Fig~\ref{fig:aextentc}) with almost the same line parameters.
The near kinematic distance corresponding to the central velocity of 8 \kms\
at  $l = $ 5\deg\ (for $l < $ 4\deg\ the estimated distance increases) is $\sim$ 2.5 kpc.
A 5\deg\ wide cloud at a distance of 2.5 kpc would have a physical size of
$\sim$ 220 pc.  This is a fairly large diffuse \CII region.  

\begin{figure}
 \resizebox{\hsize}{!}{\includegraphics{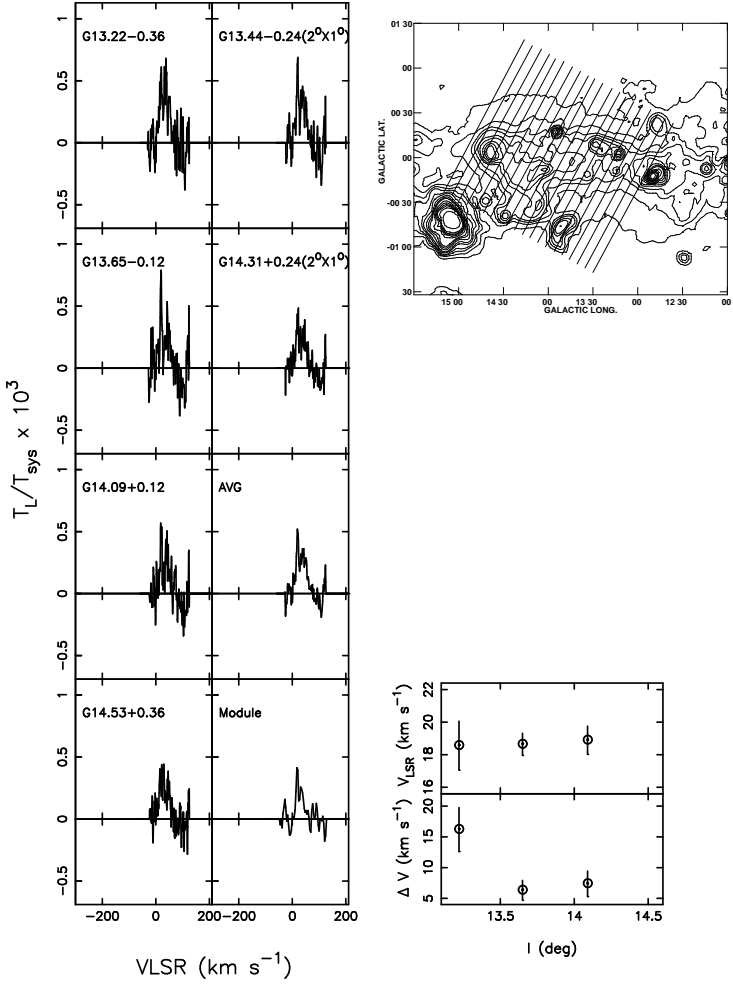}}
\caption{
Carbon line emission toward a 2\deg\ (along $l$) wide region 
centered at $l = $ 13\deg.9 and $b = $ 0\deg.
The spectra on the left are obtained by averaging the high-resolution
(2\deg $\times$ 6\arcmin) survey data
over a 2\deg $\times$ 0\deg.5 region centered at the galactic coordinates
indicated in each frame. The top two spectra
on the right are obtained by averaging the data over 
2\deg $\times$ 1\deg\ region and that labeled ``AVG'' is obtained
by averaging the data over the 2\deg\ region. The spectrum marked 
``Module'' is from the low-resolution
(2\deg $\times$ 2\deg) survey toward $l = $ 13\deg.9 and $b = $ 0\deg.  
The LSR velocity and the line width
using the spectra averaged over a 2\deg $\times$ 0\deg.5 region,
as a function of galactic longitude are shown
in the plot on the right-hand side. The vertical bars represent the 3 $\sigma$ 
error in the estimated parameters.
The narrow component
is observed in all the spectra with the same central velocity indicating the presence of
an extended diffuse \CII region.  The
observed positions are marked by the slanted lines on the 11 cm continuum
map from Reich \etal (\cite{rfrr90}). The slanted lines represent the 2\deg $\times$ 6\arcmin
beam of the ORT. 
}
\label{fig:aextentb}
\end{figure}

\clearpage
\begin{table}
\label{tab:lextentb}
\caption{Summary of the study of angular extent of carbon line emission in field 2 }
\begin{tabular}{rrrrrccrc}
\hline \hline
\multicolumn{1}{c}{$Position$} & \multicolumn{1}{c}{$T_L/T_{sys}$\footnotemark[1]} 
& \multicolumn{1}{c}{$\Delta  V$} & \multicolumn{1}{c}{$V_{\rm{LSR}}$} &  
$V_{res}$\footnotemark[3] & RMS\footnotemark[2] &\multicolumn{1}{c}{$t_{int}$}  \\
       &\multicolumn{1}{c}{$\times$ 10$^{3}$}&\multicolumn{1}{c}{(\kms)} &\multicolumn{1}{c}{(\kms)}
     & (\kms)     & $\times$ 10$^{3}$ & \multicolumn{1}{c}{(hrs)}  \\
\hline
\multicolumn{7}{c}{\bf Average over 2\deg $\times$ 1\deg} \\
\hline
    G2.25$+$0.0 & 0.35(0.04) & 14.7(1.8) &    7.9(0.7) & 2.1 & 0.07 & 95.9   \\
    G3.25$+$0.0 & 0.36(0.03) & 17.7(2.0) &    7.6(0.8) & 2.1 & 0.07 & 85.4   \\
    G4.25$+$0.0 & 0.37(0.04) & 14.0(1.7) &    8.5(0.7) & 2.1 & 0.07 & 93.7   \\
    G5.25$+$0.0 & 0.51(0.04) & 12.9(1.1) &    8.8(0.5) & 2.1 & 0.07 & 84.8   \\
    G6.25$+$0.0 & 0.36(0.05) & 11.1(1.7) &    8.2(0.7) & 2.1 & 0.08 & 64.2   \\
\hline
\multicolumn{7}{c}{\bf Average over 2\deg $\times$ 5\deg} \\
\hline
    G4.25avg    & 0.39(0.02) & 14.4(0.9) &  8.6(0.4) & 2.1 & 0.04 & 424 \\ 
\hline
\end{tabular}
\end{table}
\footnotetext[1]{The line intensities are given in units of $T_L/T_{sys}$,
where $T_L$ is the line antenna temperature and $T_{sys}$ is the
system temperature}
\footnotetext[2]{RMS is in units of $T_L/T_{sys}$.}
\footnotetext[3]{The resolution to which the spectra are
smoothed for estimating the line parameters. }

\begin{table}
\caption{Summary of the study of angular extent of carbon line emission in field 3}
\label{tab:aextenta}
\begin{tabular}{rrrrrccrc}
\hline  \hline
\multicolumn{1}{c}{$Position$} & \multicolumn{1}{c}{$T_L/T_{sys}$\footnotemark[1]} & 
\multicolumn{1}{c}{$\Delta  V$} & \multicolumn{1}{c}{$V_{\rm{LSR}}$} &  $V_{res}$\footnotemark[3] 
& {RMS\footnotemark[2]} &\multicolumn{1}{c}{$t_{int}$}  \\
  &\multicolumn{1}{c}{$\times$ 10$^{3}$}&\multicolumn{1}{c}{(\kms)} &\multicolumn{1}{c}{(\kms)}     & (\kms)     & $\times$ 10$^{3}$ & \multicolumn{1}{c}{(hrs)}   \\ 
\hline
\multicolumn{7}{c}{\bf Average over 2\deg $\times$ 0\deg.5} \\
\hline
   G13.22$-$0.36 & 0.41(0.09) &  8.7(2.1) &   50.9(0.9) & 1.8 & 0.13 & 43.6   \\
                  & 0.57(0.08) & 11.1(1.7) &   36.5(0.7) & 1.8 & 0.13 & 43.6   \\
                  & 0.50(0.06) & 16.2(2.4) &   18.6(1.0) & 1.8 & 0.13 & 43.6   \\
   G13.65$-$0.12 & 0.33(0.05) & 32.4(5.4) &   42.4(2.3) & 1.8 & 0.14 & 46.3   \\
                  & 0.73(0.11) &  6.3(1.1) &   18.6(0.5) & 1.8 & 0.14 & 46.3   \\
   G14.09$+$0.12 & 0.48(0.08) &  7.4(1.4) &   18.9(0.6) & 1.8 & 0.11 & 43.9   \\
                  & 0.30(0.04) & 22.3(3.8) &   40.9(1.6) & 1.8 & 0.11 & 43.9   \\
   G14.53$+$0.36 & 0.33(0.04) & 28.3(3.5) &   23.3(1.5) & 1.8 & 0.10 & 49.1   \\
                  & 0.24(0.05) & 12.6(3.3) &   48.5(1.4) & 1.8 & 0.10 & 49.1   \\
\hline
\multicolumn{7}{c}{\bf Average over 2\deg $\times$ 1\deg} \\
\hline
   G13.44$-$0.24 & 0.36(0.03) & 40.6(4.4) &   36.8(1.9) & 1.8 & 0.11 & 89.9   \\
                  & 0.48(0.09) &  5.9(1.3) &   18.5(0.5) & 1.8 & 0.11 & 89.9   \\
   G14.31$+$0.24 & 0.25(0.06) &  8.5(2.3) &   17.8(1.0) & 1.8 & 0.09 & 93.0   \\
                  & 0.27(0.03) & 38.5(4.5) &   35.8(1.9) & 1.8 & 0.09 & 93.0   \\
\hline
\multicolumn{7}{c}{\bf Average over 2\deg $\times$ 2\deg} \\
\hline
   G13.88$+$0.00 & 0.32(0.02) & 41.5(2.8) &   36.7(1.2) & 1.8 & 0.06 & 182.9   \\
                  & 0.35(0.04) &  6.8(1.0) &   18.4(0.4) & 1.8 & 0.06 & 182.9   \\
\hline
\end{tabular}
\end{table}
\footnotetext[1]{The line intensities are given in units of $T_L/T_{sys}$,
where $T_L$ is the line antenna temperature and $T_{sys}$ is the
system temperature}
\footnotetext[2]{RMS is in units of $T_L/T_{sys}$.}
\footnotetext[3]{The resolution to which the spectra are
smoothed for estimating the line parameters.}

\clearpage
 
\subsubsection{Carbon Line Emission toward Field 3}

Within the 2\deg\ wide field centered at $l = $ 13\deg.9, $b = $ 0\deg\ , 20 positions 
were observed with a 2\deg $\times$ 6\arcmin\ beam as shown in Fig.~\ref{fig:aextentb}.
To improve the signal-to-noise ratio on the line emission from any extended \CII\ region
we averaged the data over a 2\deg $\times$ 0\deg.5 region.  The resultant spectra are shown in the 
left hand side panels of  Fig.~\ref{fig:aextentb}.  
The carbon line in the spectra is clearly composed of a narrow and a broad component.
These spectra were further
integrated over two sets of 10 positions giving a spectrum
of a region which is 2\deg $\times$ 1\deg\ large.  These two spectra are
shown in the top two right hand side panels.  
The observed carbon line profile is well-fitted by a narrow ($\Delta V = $ 7 \kms )
and a broad ($\Delta V = $ 42 \kms) Gaussian.
Detailed line parameters obtained from the Gaussian fits are listed in Table.~\ref{tab:aextenta}.   
In the lower two right hand side panels of Fig.~\ref{fig:aextentb}, the 
high resolution spectra averaged
over a 2\deg\ region and the low resolution spectrum over the same region
are shown.  The two spectra match well within errors and
clearly show the presence of the two components.  
Since the wide component is likely to be a blend of many narrow components with
slightly different velocities, 
we require more sensitive and higher angular resolution observations to resolve
the broad component into the individual components.
The narrow component is likely to arise in a single cloud
which is at least 2\deg\ in extent along galactic longitude.  Moreover, 
the gas toward this longitude has a latitude extent of $\pm 3$ \deg\ (see Section 4) 
and the spectra toward $b \neq 0 $\deg\ shows the presence of
a narrow component (see Fig.~\ref{fig:lat2}) with almost similar line parameters
as those obtained for  
the narrow component toward this direction in the Galactic plane.
A slight increase in line width observed at higher latitudes might be a result of the poor 
signal-to-noise ratio of the spectra at
higher latitudes compared to those near $l = $ 0\deg.0, which makes the 
Gaussian decomposition of the broad and narrow features somewhat uncertain.
Thus, it appears that the diffuse \CII region in this direction is extended over
$\sim$ 6\deg\ in latitude and at least 2\deg\ in longitude.

The line-of-sight toward this longitude intercepts the spiral arms 
3, 2 and 4 which are nominally located at radial distances of 
$\sim$ 1.9, 3.7 and 14.1 kpc respectively from the 
Sun (see Fig.~\ref{fig:galdis}).  The near and far 
kinematic distances corresponding to the observed central velocity (18.4 \kms) of the 
narrow component are 2.3 and 14.2 kpc.
If the cloud is located at the near distance and  
the angular extent of the narrow line emitting region
is at least 2\deg\ $\times$ 6\deg\ then it corresponds to a physical size perpendicular
to the line-of-sight  $> 80 \times 200$ pc.  This, again, is a fairly large diffuse \CII
region.

In summary, our data toward Field 2 and 3 indicate the presence of 
extended \CII\ regions - extending over $\sim$ 200 pc or more. 
Line emission from many other positions suggests that structure in diffuse \CII regions
on scales of $\sim$ 6 pc is common.   
 
\section{Summary}
\label{sec:sum}

The paper discusses the carbon line emission detected in the low-resolution
(Paper I) and the high-resolution surveys (Paper II) of recombination lines near 327 MHz made 
in the galactic plane using the ORT. The observed carbon line parameters in the
high-resolution survey are presented. The carbon lines observed in the surveys arise
in diffuse \CII regions unlike the high frequency ($ \nu > 1$ GHz) recombination lines which
arise in the photo-dissociation regions associated with \HII regions.  
Most of our carbon line detections are in the longitude range $ l = 358$\deg\ $\rightarrow$ 20\deg\
with a few detections between $l = 20$\deg\ to 89\deg.
At longitudes $l = $ 0\deg\ and 13\deg.9, observations
with a 2\deg\ $\times$ 2\deg\ beam along galactic latitude indicate that line emission extends
at least up to $b \sim \pm$ 3\deg.
 
The \lv diagram and radial distribution constructed using
the carbon line data near 327 MHz show similarity with those obtained from
the hydrogen RRLs near 3 cm from \HII regions.  The radial extent of 
carbon line emission also resembles that of ``intense'' \CO emission.  These similarities
suggest that the distribution of the carbon line forming regions resembles the
distribution of star-forming regions in our Galaxy. 
The \lv diagram of carbon line emission near 327 MHz
is similar to those obtained from the
carbon lines detected in absorption at frequencies near 76 MHz (Erickson \etal \cite{ema95}) 
and 35 MHz (Kantharia \& Anantharamaiah \cite{ka01}), indicating that the latter
are the absorption counterparts of the carbon lines observed in emission
near 327 MHz.

We estimated the \FIRCII intensity expected from low frequency 
carbon RRL forming regions coexistent with the CNM and low-density PDRs.
The estimate was made using a subset of physical parameters ($T_e$ = 
$20 \rightarrow 80$ K) determined by Kantharia \& Anantharamaiah (\cite{ka01}) for the
diffuse \CII regions.
Significant fraction  ($\sim$ 95 \%) of the observed FIR line emission can arise in CNM/low-density PDR if
the temperature of the diffuse \CII region is close to $\sim$ 80 K whereas only a small
fraction (0.4 \%) of the observed FIR line emission can be produced in regions with temperature near 20 K.
We then compared the
longitudinal distribution of the two tracers of ionized carbon in the inner Galaxy
to investigate their common origin.  
Available data do not rule out the possibility that
diffuse \CII regions contribute significantly to the FIR line emission in the inner Galaxy.  
However, data with comparable angular and spectral resolutions are required to further
investigate this possibility.

Analysis of our high resolution data shows that the diffuse \CII regions exhibit
structure over angular scales ranging from $\sim 6$\arcmin\ to $\sim 5$\deg.
Toward $l=35$\deg and several other directions, the diffuse \CII regions
exhibit structure over
$\sim 6$ \arcmin, manifested as different central line velocities and line widths.
At $l = $ 35\deg\, we estimated a linear size of $\sim 6$ pc
for a clump when placed at the near kinematic
distance.  Thus, it is likely that there exist \CII regions of this size (or smaller) or there
exist clumps of this size embedded in a larger diffuse \CII region.  
Toward $l = $ 13\deg.9, a narrow ($\sim$ 7 \kms) carbon line emitting region with an
angular extent $\ge 2$\deg\ in galactic longitude and $\sim \pm$ 3\deg\ in galactic latitude
has been observed. 
A similar extended diffuse \CII region was identified toward 
$l = $ 4\deg.25, $b= $ 0\deg.
The angular extent of this region is at least 5\deg\ along $l$ and 2\deg\ along $b$.
Such large angular sizes translate to physical sizes perpendicular to the sight-lines
of $>$ 200 pc.  These are fairly large diffuse \CII regions.  
Thus, our data shows 
that the diffuse \CII regions observed in the inner Galaxy 
display structure on scales ranging from a few parsecs to a couple of hundred parsecs.  

The association of some of the observed carbon line emission near 327 MHz in the surveys
with \HI self-absorption features will be discussed in 
Roshi, Kantharia \& Anantharamaiah (\cite{rka02}). 

\begin{acknowledgements}

DAR thanks F. J. Lockman and D. S. Balser for the many stimulating discussions and helpful
suggestions during the course of this work.  NGK thanks Rajaram Nityananda for useful suggestions.
We are grateful to late K. R. Anantharamaiah for
his guidance and support, to which we owe much of the work we have
accomplished in our careers.  
We thank the referee F. Wyrowski for his comments and suggestions which
have improved the paper.

\end{acknowledgements}

\end{document}